\documentclass[11pt]{article}

\usepackage{amssymb,amsmath,amsthm}
\usepackage{bbm}
\usepackage[noend]{algpseudocode}
\usepackage{graphicx}
\usepackage{verbatim}
\usepackage{url,xspace,hyperref}
\usepackage{cite}
\usepackage{caption}
\usepackage{subcaption}
\usepackage{multirow}
\usepackage{multicol}
\usepackage{latexsym}
\usepackage{amsmath,amssymb,enumerate}
\usepackage{xspace}
\usepackage{algorithm,algorithmicx}
\usepackage{float}
\usepackage{xcolor}
\usepackage{mathrsfs}
\usepackage{cleveref}

\usepackage{fullpage}
\usepackage{geometry}
\usepackage{setspace}

\newtheorem{theorem}{Theorem}[section]
\newtheorem*{theorem*}{Theorem}

\newtheorem{lemma}[theorem]{Lemma}

\newcounter{note}[section]

\def\sse{\subseteq}

\newcommand{\pr}{\mathbf{P}} 
\newcommand{\E}{\mathbb{E}} 
\newcommand{\ZZ}{\mathbb{Z}}

\newcommand{\R}{\mathbb{R}}
\newcommand{\cR}{\mathcal{R}}
\newcommand{\A}{\mathcal{A}}

\newcommand{\OPT}{\mathtt{OPT}}
\newcommand{\NA}{\mathtt{NA}}

\newcommand{\AD}{\mathtt{AD}}

\newcommand{\exsfe}{$\mathtt{EX}$-d-$\mathtt{SHE}$\xspace}
\newcommand{\she}{$\mathtt{SHE}$\xspace}

\newcommand{\ignore}[1]{}
\newcommand{\rv}[1]{\mathtt{{#1}}}

\newcommand{\ssc}{$\mathtt{SSClass}$\xspace}

\newcommand{\sck}{$\mathtt{StocCovKnap}$\xspace}

\newcommand{\bX}{X}

\newcommand{\NAC}{\textsc{NaCl}\xspace}

\newcommand{\ci}{\ell}
\renewcommand{\endproof}{\hfill$\square$}

\title{Non-Adaptive Stochastic Score Classification and \\Explainable Halfspace Evaluation}
\author{Rohan Ghuge\thanks{Department of Industrial and Operations Engineering, University of Michigan, Ann Arbor, USA. Research supported in part by NSF grants CMMI-1940766 and CCF-2006778.} \and Anupam Gupta\thanks{Department of Computer Science, Carnegie Mellon University, Pittsburgh, USA. Supported in part by NSF awards CCF-1907820, CCF-1955785, and CCF-2006953.} \and Viswanath Nagarajan$^*$}

\begin{document}
\maketitle

\begin{abstract}
Sequential testing problems involve a complex system with several  components, each of which is ``working"  with some independent probability. 
The outcome of each component can be determined by performing a test, which incurs some cost. 
The overall system status is given by  a function $f$ of the outcomes of its components.  
The goal is to  evaluate this function $f$ by performing tests at the minimum expected cost.  While there has been extensive prior work on this topic, provable approximation bounds are mainly limited to  simple functions like ``k-out-of-n'' and halfspaces.  
We consider significantly  more general ``score classification" functions, and provide the first constant-factor approximation algorithm (improving over a previous logarithmic approximation ratio). Moreover, our policy is non adaptive: 
it just involves performing tests in an {\em a priori} fixed order.
We also consider  the  related 
halfspace evaluation problem,  where we want to evaluate some function on  $d$ halfspaces (e.g., intersection of halfspaces).   We show that our approach provides an $O(d^2\log d)$-approximation algorithm for  this problem.   Our algorithms also extend  to the setting of ``batched'' tests, where multiple tests can be perfomed simultaneously while incurring an extra setup cost. 
Finally, we perform computational experiments that demonstrate the practical performance of our algorithm for score classification. 
We observe that, for most instances, the cost of our algorithm is within $50\%$ of an information-theoretic  lower bound on the optimal value.
\end{abstract}

\maketitle

\section{Introduction}

The problem of diagnosing complex systems often involves running costly tests for  several components of such a system. 
One approach to diagnose such systems is to perform tests on all components, which can be prohibitively 
expensive. A better and more practical approach involves \emph{sequential testing}, where  a policy  tests components one by one until the system is diagnosed, while  minimizing the expected  cost of testing. Such sequential testing problems arise in a number of applications  such as healthcare, manufacturing and telecommunucation; we refer to the surveys by \cite{Unluyurt04} and  \cite{Moret82} for more details.

Concretely, there are $n$ components, where each 
component is ``working'' independently with 
some probability. The outcome (working or failed) of each component can be determined by performing a test, which incurs some cost. 
The overall status of the system is given by a function $f:\{0,1\}^n\rightarrow \ZZ_+$ of the outcomes of its $n$ components. The goal is to determine $f(X_1,\cdots X_n)$, where $X_i$ denotes the outcome of component $i$. The objective is to minimize the expected cost of testing.   

Perhaps the simplest and most commonly studied function is the AND-function (or series system), which involves testing if all the components are working, i.e., $\sum_{i=1}^n X_i=n$. In this case, it is well known that the natural greedy algorithm finds an optimal solution, see e.g.,~\cite{B72}. An exact algorithm is also known for $k$-out-of-$n$ functions, where we want to determine if at least a threshold number $k$ of components are working (i.e., $\sum_{i=1}^n X_i\ge k$); this result is by \cite{Ben-Dov81}.  More generally, in the halfspace evaluation problem, each component $i$ has a weight $w_i$ and we want to check whether the total weight $\sum_{i=1}^n w_i\cdot X_i$ of working components is above/below a threshold. \cite{DeshpandeHK16} obtained a constant-factor approximation algorithm for the halfspace evaluation problem (it is also known to be NP-hard).

In this paper, we obtain sequential testing algorithms with provable guarantees for much more general functions $f$. In particular, we consider the stochastic score classification problem, introduced by \cite{GkenosisGHK18}, where components have weights as in the halfspace evaluation problem. 
We are also given multiple threshold values 
that partition the number line into $B$ disjoint intervals, 
and the goal is to identify the interval that the total weight $\sum_{i=1}^n w_i\cdot X_i$ lies in. For example, the intervals may correspond to the overall system status being poor/fair/good/excellent, and we want to identify the system status at minimum expected cost. 
Note that the halfspace evaluation problem is a special case of score classification when there are just $B=2$ intervals. 

We also consider the problem of evaluating the {\em intersection} of $d$ halfspaces. Here, there are $d$ different halfspaces and we want to check whether all the halfspaces are satisfied (and if not, we also want to identify some violated halfspace). In fact, we consider a more general problem that involves evaluating an arbitrary function over the $d$ halfspaces.

Solutions to all these problems 
are \emph{sequential decision processes}, where one component  is tested  at each
step (after which its
outcome is observed). This
process continues until the function value is determined. We note that solutions in general are \emph{adaptive}: the choice of the next component  to test may depend on all previous outcomes. In practice, it is often preferable to use   a \emph{non-adaptive} solution, which is simply given by a
priority list of components:  tests are then performed in that {\em fixed} order until the function is evaluated.
Non-adaptive  solutions are 
simpler and faster to implement (compared to their adaptive counterparts) as the  testing sequence needs to be constructed just once, after which it can be used for all  input realizations. 
However, non-adaptive solutions typically cost more than adaptive ones. So, one would ideally like to find a non-adaptive solution of cost comparable to an optimal adaptive solution. A priori, it is not even clear if such a low cost non-adaptive solution exists (let alone finding it in polynomial time). The \emph{adaptivity gap}, introduced by \cite{DGV08}, 
captures the worst-case ratio between the best non-adaptive and best adaptive solutions; note that this quantity is independent of computational efficiency.

  
Our main result is a non-adaptive algorithm for stochastic score classification that is a constant-factor approximation to the optimal adaptive solution. We also obtain a non-adaptive $O(d^2\ln d)$ factor approximation algorithm for evaluating the intersection of $d$ halfspaces. Both these results involve efficient algorithms (in fact, nearly linear time) and also bound the adaptivity gaps. 

We also consider a more general cost-structure, where multiple tests can be performed simultaneously while incurring an extra set-up cost.  As discussed in \cite{SegevShaposhnik22}, this feature is present in many practical applications as it allows batching tests to achieve economies of scale. Our results extend  to this batch-cost setting as well.

\subsection{Problem Definitions}

In the following, for any integer $m$, we use   $[m] := \{1,2,\ldots, m\}$.

\paragraph{Stochastic Score Classification (\ssc).} An instance consists of $n$ tests (a.k.a. items). Each item $i\in [n]$ is associated with an 
independent $\{0,1\}$ random
variable (r.v.) $X_i$ with $\Pr[X_i = 1] = \E[X_i] = p_i$. We use 
 $\bX = \langle X_1, \ldots, X_n\rangle$ to denote the vector of all r.v.s. The algorithm knows the probability values $p_i$s in advance, but not the random outcomes $X_i$s. In order to determine the outcome $X_i$ of item $i$, the algorithm needs to 
 probe/test $i$ by incurring a cost $c_i\in \R_+$. We are also given 
 weights $w_i\in \ZZ$ for all items $i\in [n]$, and the
total weight of any outcome $\bX$ is denoted
$w(\bX) = \sum_{i=1}^n w_i X_i$. In addition, we are given $B+1$
integer thresholds $\alpha_1 < \alpha_2 <  \ldots < \alpha_{B+1}$ such that \emph{class} $j$
corresponds to the interval
$I_j := \{\alpha_j, \ldots, \alpha_{j+1} - 1\}$. We assume that $\alpha_1$ (resp. $\alpha_{B+1}$) is the minimum (resp. maximum) total weight of any outcome. So, the intervals $I_1,\cdots I_B$ form a partition of possible weight outcomes. We want to evaluate the \emph{score
  classification function} $f: \{0, 1\}^n \to \{1, \ldots, B\}$ where $f(\bX)=j$ precisely when $\alpha_j \le w(\bX) < \alpha_{j+1}$, i.e. $w(\bX) \in I_j$. The goal is to
determine $f(\bX)$ at minimum expected cost for testing.

Let $W:=\sum_{i=1}^n w_i$ denote the total weight of all items. For each class $j\in [B]$, we  associate two numbers
$\beta_j^0 = W - \alpha_{j+1}+1$ and $\beta_j^1 = \alpha_j$. Note that for any outcome $X$ and class $j$,
\begin{equation}\label{eq:ssc-f-eval}
f(X)=j\mbox{ if and only if  } w(X)\ge \beta^1_j \mbox{ and }W-w(X)\ge \beta^0_j.
\end{equation}

In  Appendix~\ref{app:positive-wt}, we show that any instance of \ssc with
both positive and negative weights can be reduced to an equivalent instance
with all positive weights. 
Henceforth,  we  
assume (without loss of generality) that all weights $w_i$s are non-negative. In this case, we also assume that $\alpha_1=0$ and $\alpha_{B+1}=W$. 

In our algorithm, it is convenient to introduce two (nonnegative) random rewards $R_0(i) := w_i\cdot (1-X_i)$ and $R_1(i) := w_i \cdot X_i$ associated with each item $i\in [n]$. Note that the total $R_1$-reward (resp. $R_0$-reward) corresponds to the total weight of working (resp. failed) items.  For any subset $S\sse [n]$ of items, we define $R_0(S)=\sum_{i\in S} R_0(i)$ and
$R_1(S)=\sum_{i\in S} R_1(i)$.

  \paragraph{Explainable Stochastic $d$-Halfspace Evaluation (\exsfe).}
An instance consists of $n$ random items $\bX = \langle X_1, \ldots, X_n\rangle$   as in \ssc, and 
$d$  halfspaces. For each $k\in [d]$, the $k^{th}$ halfspace  $h_k(X)$  is given by  $\sum_{i=1}^n w_{ki}X_i \ge \alpha_k$; we set $h_k(X)=1$ if the halfspace is satisfied and $h_k(X)=0$ otherwise.   Moreover, we are also given an \emph{aggregation
    function} $f: \{0,1\}^d \to \{0,1\}$, and we want to evaluate
  $f(h_1(X), \ldots, h_d(X))$. In other words, if we define the composite  function $h(X) := (h_1(X), h_2(X), \ldots, h_d(X))$,
  we want to evaluate $f \circ h(X)$.  
Furthermore, the problem asks for an ``explanation'', or a {\em witness} of
  this evaluation. The definition of a witness is a bit technical, so it is deferred to \S\ref{sec:general}.  
 The goal is to evaluate $f\circ h$ along with a witness, at the minimum expected
  cost.

For example, when $f(y_1,\cdots y_d) = \bigwedge_{k=1}^d y_k$, \exsfe\ corresponds to evaluating  the intersection of $d$  halfspaces; the witness  in this case either confirms that all halfspaces are satisfied or identifies some violated halfspace. 
\exsfe\ also captures more general cases: given a target $\bar{d}\le d$, we can  check whether at least $\bar{d}$ out of $d$ halfspaces are satisfied (using an appropriate $f$).

\paragraph{Batch Cost Structure.}  This is a generalization of the basic (additive) cost structure, where any subset of items may be tested simultaneously by incurring an extra setup cost $\rho$. Formally, the cost to simultaneously test a subset $S\sse [n]$ of items  is $\rho + \sum_{i\in S} c_i$. Note that setting $\rho=0$, we recover the usual cost structure (as in \ssc and \exsfe), in which case there is no benefit to batching tests together. However, when $\rho$ is large, it is beneficial to perform multiple tests simultaneously because one can avoid paying the setup cost repeatedly. Here, a solution involves selecting a batch  of items in each step (instead of a single item). Again, the goal is to evaluate the function (in \ssc or \exsfe) at minimum expected cost.

\subsection{Results and Techniques}
\label{subsec:results}

Our main result is the following algorithm (and adaptivity gap). 
\begin{theorem} \label{thm:main}
    There is a  
    non-adaptive algorithm for stochastic score classification with expected cost at most a constant factor times the optimal adaptive cost.
\end{theorem}

This result improves on the prior
work from~\cite{GkenosisGHK18} in several ways. Firstly,
we get a constant-factor approximation, improving upon the previous
$O(\log W)$ and $O(B)$ ratios, where $W$ is the sum of weights, and
$B$ the number of classes. Secondly, our algorithm is non-adaptive in
contrast to the previous adaptive ones. Finally, our algorithm has
nearly-linear runtime, which is faster than the previous
algorithms. 

An added benefit of our approach is that we obtain a ``universal''
solution that is simultaneously $O(1)$-approximate for all
class-partitions. Indeed, the non-adaptive list produced by our algorithm only depends on the probabilities, costs and weights,
and not on the class boundaries $\{\alpha_j\}$; these $\alpha_j$
values are only needed in the stopping condition.

Our second result is the following algorithm for \exsfe.
\begin{theorem}\label{thm:exsfe-main}
There is a  
    non-adaptive algorithm for explainable stochastic $d$-halfspace evaluation with expected cost $O(d^2\log d)$ times the optimal adaptive cost.
\end{theorem}
As a special case, we obtain a non-adaptive $O(d^2\log d)$-approximation algorithm for (explainable) intersection of $d$ halfspaces.
 The stochastic intersection of halfspaces problem (in a slightly different model) was studied previously by  \cite{BlancLT21}, where the solution may make errors with probability $\delta>0$.  Assuming  that all probabilities $p_i=\frac12$, \cite{BlancLT21} obtained  an $O\left(\sqrt{n \log d}/{\delta^2}\right)$-approximation algorithm. Another difference from our model is that \cite{BlancLT21} do not require a witness. So their policy can stop if it  concludes that there exists a  violated halfspace (even without knowing which one), whereas our policy can only stop after it identifies a violated halfspace (or determines that all halfspaces are satisfied). We note that our approximation ratio is 
 independent of the number of variables $n$ and holds for arbitrary probabilities. 

Next, we consider the more general batch-cost structure and show the following. 
\begin{theorem}\label{thm:batched-ssc-main}
In the setting of batch-costs, there is  a non-adaptive algorithm with:
\begin{itemize}
\item approximation ratio $O(1)$ for stochastic score classification.
\item approximation ratio $O(d^2\log d)$ for stochastic $d$-halfspace evaluation.
\end{itemize}
\end{theorem}
To the best of our knowledge, this is the first constant approximation even for halfspace evaluation 
 in the batched setting. Previously, \cite{DaldalGS+16} and \cite{SegevShaposhnik22} obtained constant-factor approximation algorithms for evaluating an AND-function in the batched setting. 


Finally, we evaluate the empirical performance of our algorithm for score classification. In these experiments, our non-adaptive algorithm performs nearly as well as
the previous-best \emph{adaptive} algorithms, while
being an order of magnitude faster.  In fact, on many instances, our
algorithm provides 
an \emph{improvement} in both the cost as well as the running time. On most instances, the cost of our algorithm is within $50\%$ of an information-theoretic  lower bound on the optimal value.

\begin{figure}
     \centering
     \begin{subfigure}[b]{0.45\textwidth}
         \centering
         \includegraphics[width=3in]{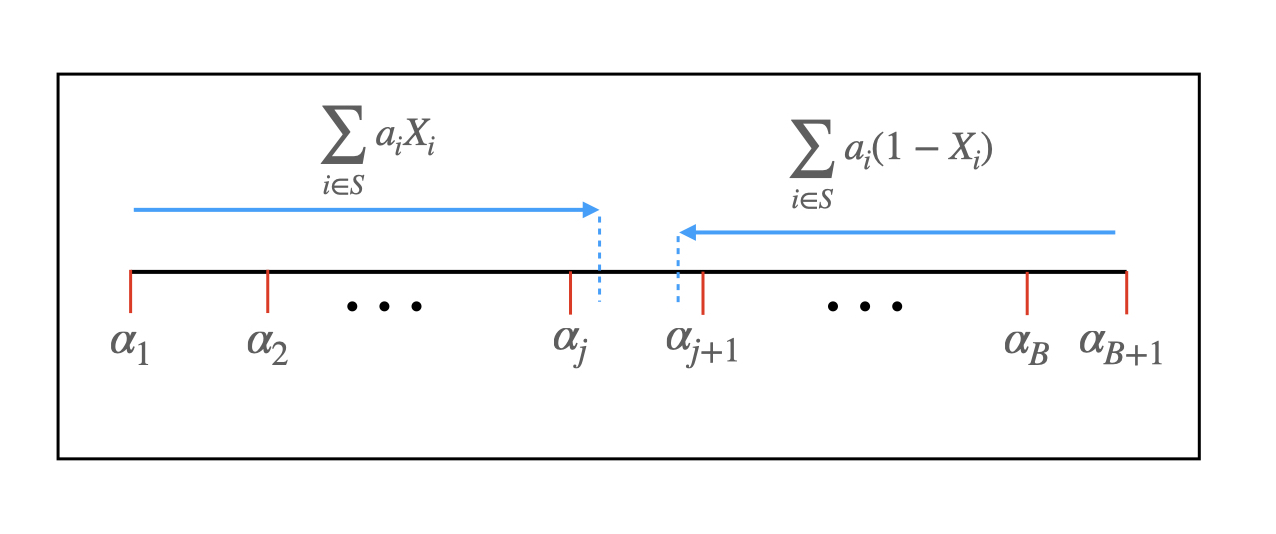}
         \caption{In this case, condition \eqref{eq:ssc-stop-cond} is satisfied for class $j$ and testing can stop.}
     \end{subfigure}
     \quad
     \begin{subfigure}[b]{0.45\textwidth}
         \centering
         \includegraphics[width=3in]{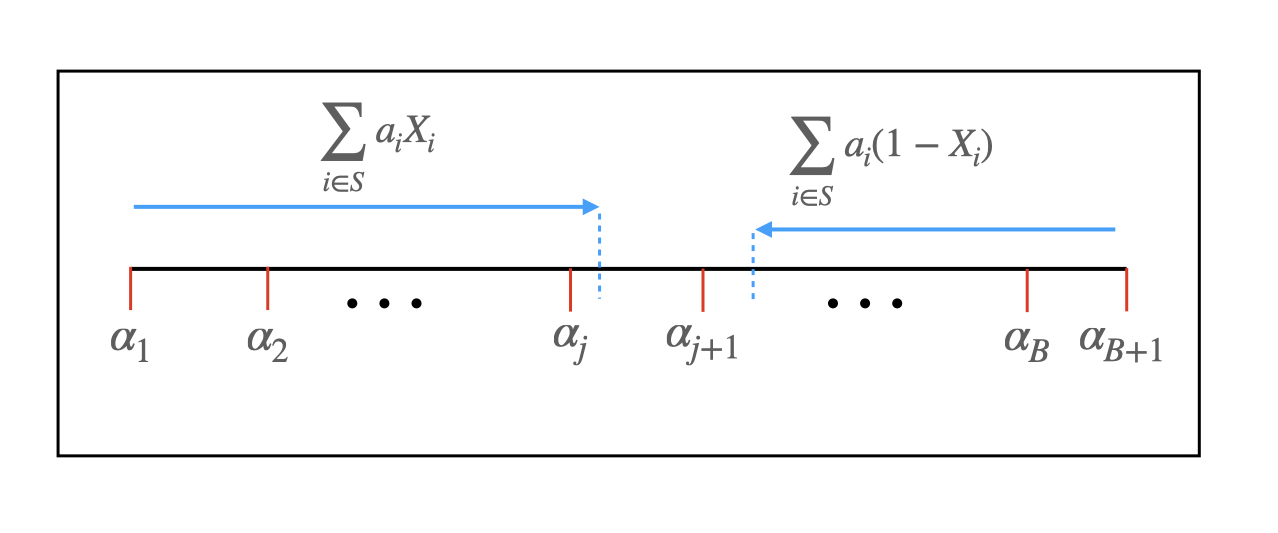}
         \caption{Here, condition \eqref{eq:ssc-stop-cond} is not satisfied:  $f(\bX)$ could be $j$ or $j+1$, so testing must continue. }
     \end{subfigure}
     \caption{Illustration of our approach for \ssc.}
        \label{fig:probing}
\end{figure}

\paragraph{\bf Overview of techniques.} 
To motivate our algorithm for \ssc, suppose that we have currently tested/probed a subset
$S\sse [n]$ of items. Then, we can evaluate the score-classification function $f$ if and only if, there is some class $j\in[B]$  
such that
\begin{equation}\label{eq:ssc-stop-cond}
\sum_{i \in S} w_i X_i \geq \beta_j^1 = \alpha_j\quad \mbox{ and }\quad
 \sum_{i\in S} w_i(1-X_i) \geq \beta_j^0 = W - \alpha_{j+1}+1.
 \end{equation}
 Indeed, if the above condition is satisfied then, irrespective of the outcomes of the untested items $[n] \setminus S$, we have $w(X)\ge  \sum_{i \in S} w_i X_i \geq \beta_j^1$ and $W-w(X)\ge  \sum_{i\in S} w_i(1-X_i) \geq \beta_j^0  $; this uses the assumption that all weights are nonnegative. Using \eqref{eq:ssc-f-eval}, we can then 
conclude that the function value $f(X)=j$.   On the other hand, if the above condition is
not satisfied for {\em any } class $j$, we must continue testing as the function value cannot be determined yet. See 
\Cref{fig:probing} for an example. Note that  condition \eqref{eq:ssc-stop-cond} is the same as $R_0(S)\ge \beta_j^0$ and $R_1(S)\ge \beta_j^1$; recall the definition of the $R_0$ and $R_1$ rewards. 
 Our algorithm tries to achieve condition \eqref{eq:ssc-stop-cond} at the minimum testing cost.

The challenge however is that the class $j$ (of the realization $X$) is unknown: so we do not have a numeric threshold on the $R_0$ or $R_1$ rewards to aim for. If we knew the target $R_0$ and $R_1$ rewards then we could directly use constant-factor approximation algorithms for the stochastic covering knapsack problem from \cite{DeshpandeHK16} or \cite{JiangLL+20}. In the {\em stochastic covering knapsack} (\sck) problem, given items with deterministic costs and random rewards, we want to minimize the expected cost to achieve a target total reward value. This is precisely what we would want to solve in \eqref{eq:ssc-stop-cond} if the class $j$ was fixed.



We get around this issue (of not knowing class $j$) by coming up with a ``universal'' solution for \sck that works for {\em all} targets simultaneously. Roughly speaking, we build two universal solutions $L_0$ and $L_1$, corresponding to covering the $R_0$ and $R_1$ rewards respectively. Then, we interleave these two solutions so that the cost incurred in the combined solution is split evenly between $L_0$ and $L_1$. The key step in building this universal solution lies in coming up with a non-adaptive solution for \sck with the following guarantee:  given any cost-budget $B$ and tolerance $\epsilon>0$,  there is a subset $S$ of items with cost $O_\epsilon(B)$ such that for {\em every} adaptive solution $\AD$ of cost $B$, the
 probability that $S$ has more reward than $\AD$ is at least $1-\epsilon$. Our algorithm here relies on solving a logarithmic number of {\em deterministic} knapsack instances, where we use truncated expectations as the deterministic rewards. In order to choose the correct truncation threshold, we use the ``critical scaling'' idea from \cite{JiangLL+20}. We note however that the details of our 
algorithm/analysis are different: we use properties of the fractional knapsack problem and our analysis  doesn't require martingale concentration inequalities as in \cite{JiangLL+20}.


\subsection{Related Work}
Sequential testing problems have been extensively studied in operations research and computer science; see e.g., surveys by \cite{Moret82} and \cite{Unluyurt04}. 
When the function $f$ is an AND-function (equivalently, an OR-function), it is easy to easy to see that an optimal solution is non-adaptive. Moreover, \cite{B72} proved that testing items in a greedy order is optimal in this case. For $k$-out-of-$n$ functions, non-adaptive solutions are no longer optimal, and  an optimal adaptive algorithm was obtained by \cite{Ben-Dov81}. A non-adaptive $2$-approximation algorithm also follows from the work of \cite{GkenosisGHK18}. The more general halfspace evaluation (a.k.a. linear threshold functions) is known to be NP-hard, and an adaptive $3$-approximation algorithm was obtained in \cite{DeshpandeHK16}. A non-adaptive  $O(1)$-approximation algorithm for halfspace evaluation can be inferred from \cite{JiangLL+20}. 

Approximation algorithms are also known for evaluating certain other functions. In particular, \cite{AllenHK+15}  obtained a $(2k)$-approximation for monotone functions in ``disjunctive normal form'' with $k$ terms, and \cite{KaplanKM05} obtained a logarithmic approximation ratio for functions with {\em both} disjunctive  and conjunctive normal form  representations.

The \ssc problem was introduced by
\cite{GkenosisGHK18}, who showed that  it  can be formulated as an instance of ``stochastic submodular cover''. Then, using more general results (\cite{INZ12,GolovinK-arxiv}), they obtained an adaptive $O(\log W)$-approximation algorithm. Furthermore, \cite{GkenosisGHK18} obtained an adaptive  $3(B-1)$-approximation algorithm for \ssc by extending the approach of \cite{DeshpandeHK16} for halfspace evaluation. 
\cite{GkenosisGHK18} also studied the {\em unweighted} special case of \ssc (where all weights $w_i=1$) and gave
a slightly better $(B-1)$-approximation algorithm. 
A main open question from this work was the possibility of a constant approximation for the general \ssc problem. We answer this in the affirmative. Moreover, our algorithm is non-adaptive: so we also bound the adaptivity gap.

Stochastic covering knapsack (\sck) is closely related to halfspace evaluation. Indeed, the results on halfspace evaluation in \cite{DeshpandeHK16} and \cite{JiangLL+20} are based on this relation. 
We note that the results in \cite{JiangLL+20}  applied to the more general stochastic $k$-TSP problem (\cite{EneNS17}). As noted earlier, we make use of some ideas from \cite{JiangLL+20}. However, our algorithm/analysis is simpler and we obtain a nearly-linear time algorithm.  

\cite{DaldalOS+17} introduced sequential testing with {\it batch costs} and developed efficient heuristics (without performance guarantees).  Subsequently, \cite{DaldalGS+16} obtained a constant-factor  approximation algorithm for evaluating AND-functions under batch costs. Recently, \cite{SegevShaposhnik22} improved this result to a polynomial-time approximation scheme (PTAS).  We are not aware of results for more general functions (even $k$-out-of-$n$) under batch costs. 

More generally, non-adaptive solutions (and adaptivity gaps) have been used    for  various other stochastic optimization problems such as max-knapsack~(\cite{DGV08,BGK11}), matching~(\cite{BGLMNR12,BehnezhadDH20}), matroid intersection~(\cite{GN13,GuptaNS17}) and orienteering~(\cite{GuhaM09,GuptaKNR15,BansalN15}). Our result shows that this approach is  also useful for \ssc.

\paragraph{Subsequent works.} 
After the preliminary version of this paper appeared, there have been some improved results on the {\em unweighted} special case of \ssc (where all weights are one).
\cite{PlankSchewior22-arxiv} and \cite{Liu22-arxiv} obtained approximation ratios of $5.828$ and $6$, respectively.
For the further special case with unit costs, \cite{GrammelHKL22} obtained a  $2$-approximation algorithm. All these algorithms also find non-adaptive solutions. We note our results are more general  because they can handle arbitrary weights (as in halfspace evaluation). 

\subsection{Organization.}
We start with some basic results on deterministic knapsack in \S\ref{sec:prelim}. In \S\ref{sec:stoch-knap}, we present the non-adaptive algorithm for   \sck with a  strong probabilistic guarantee. This result is then used to obtain all our algorithms: \S\ref{sec:ssclass-alg} is on \ssc, \S\ref{sec:general} is on \exsfe and \S\ref{sec:batched-ssclass} is on the batch-cost extensions. 
Finally, we present  computational results in \S\ref{sec:computations}.

\section{Preliminaries on Deterministic Knapsack}\label{sec:prelim}

We first state some basic results for the deterministic knapsack problem. In an instance of the knapsack problem, we are given a set $T$ of items with non-negative costs $\{c_i : i\in T\}$ and rewards $\{r_i : i\in T\}$, and a budget $D$ on the total cost. The goal is to select a subset of items of total cost at most $D$ that maximizes the total reward. The LP relaxation is the following:
$$g(D) = \textstyle \max \left\{ \sum_{i\in T} r_i\cdot x_i \,\, \big| \,\,\sum_{i\in T} c_i\cdot x_i \le D, \, x\in [0,1]^T\right\},\qquad \forall D\ge 0.$$ 

The following algorithm $\A_{\rv{KS}}$ solves the {\em fractional} knapsack problem and also obtains an approximate integral solution. 
Assume that the items are ordered so that $\frac{r_1}{c_1}\ge \frac{r_2}{c_2}\ge \cdots$. Let $t$ index the first item (if any) so that $\sum_{i=1}^t c_i\ge D$. Let $\psi:=\frac{1}{c_t}(D-\sum_{i=1}^{t-1} c_i)$ which lies in $(0,1]$. Define
$$x_i=\left\{ \begin{array}{ll}
1 & \mbox{ if } i\le t-1\\
\psi & \mbox{ if } i=t \\
0 & \mbox{ if } i \ge t+1 \\
\end{array}\right..$$
Return $x$ as the optimal fractional solution and $Q=\{1,\cdots, t\}$ as an integer solution. 
We prove the following fact in the appendix for completeness.

\begin{theorem}\label{thm:knap}
Consider algorithm $\A_{\rv{KS}}$ on any instance of the knapsack problem with budget $D$. 
\begin{enumerate}
\item $\langle r,
  x\rangle = \sum_{i=1}^{t-1} r_i +\psi\cdot r_t =g(D)  $ and so $x$ is an optimal LP solution. 
\item The derivative $g'(D) = \frac{r_t}{c_t}$.
\item Solution $Q$ has cost $c(Q)\le D+c_{max}$ and reward $r(Q)\ge g(D)$. 
\item $g(D)$ is a concave function of $D$.
\end{enumerate}
\end{theorem}

\newcommand{\SKP}{\textsc{StochKnap}\xspace}

\section{Non-Adaptive Stochastic Covering Knapsack with Probabilistic Guarantee}\label{sec:stoch-knap}
Consider any instance of \sck with  $n$ items with deterministic costs $\{c(i)\}_{i=1}^n$ and (independent) random rewards $\{R(i)\}_{i=1}^n$. An {\em adaptive policy} is a sequence of items, where the choice of the $i^{th}$ item depends on the observed rewards on the first $i-1$ items (as well as internal random outcomes). In contrast, a {\em non-adaptive
  policy} is simply a static  sequence of the items. For any policy $\Pi$,  let $\langle \pi_1, \pi_2,\cdots \rangle$ denote  the corresponding sequence, where $\pi_i\in [n]$ is the $i^{th}$ selected item. Note that $\pi_i$s are random indices for an adaptive policy $\Pi$, whereas they are deterministic for a non-adaptive policy. We note that an adaptive policy may also use internal randomness in choosing items, i.e., the first $i-1$ items $\{ \pi_j \}_{j=1}^{i-1}$ along with their observed rewards $\{R(\pi_j)\}_{j=1}^{i-1}$ determine  the probability distribution of the next item $\pi_i$. 
    We use $R(\Pi)$ to denote the random total reward collected by policy $\Pi$. We say that policy $\Pi$ {\em costs  at most $B$} if the total cost of the (possibly random) items selected by $\Pi$ is always at most $B$. 
    The main result of this section is a non-adaptive policy that achieves 
 the following probabilistic guarantee relative to {\em any} adaptive policy.

\begin{theorem}\label{thm:stoch-knap-prob}   
Given a  set of items $N$ with costs $\{c(i)\}_{i\in N}$, random rewards $\{R(i)\}_{i\in N} $,  budget $B$, and  $\epsilon>0$,   Algorithm~\ref{alg:stoch-knap} returns a non-adaptive (i.e., fixed) subset $S\sse N$ such that
\begin{itemize}
\item  {\bf Cost guarantee:}  $c(S)\le O(\frac1\epsilon \, \ln \frac1\epsilon) \cdot B$, and 
\item {\bf Reward guarantee:} $\Pr[R(S) < R(\AD)] \le \epsilon$  for  every adaptive policy $\AD$ of cost at most $B$.
\end{itemize}
Moreover, the algorithm runs in time nearly-linear in $|N|$. 
 \end{theorem}
We note that this result is stronger than typical ``adaptivity gap'' results which only bound the {\em expected} objective values of adaptive/nonadaptive policies. The fact that we have a probabilistic guarantee (rather than just expectation) turns out to be crucial in designing the algorithms for \ssc and \exsfe. 

Our algorithm is based on solving a {\em deterministic} knapsack instance using the natural greedy algorithm $\A_{\rv{KS}}$. 
A key challenge in constructing the deterministic knapsack instance  is the choice of {\em deterministic} rewards.
To this end, we consider truncated expectations of the following form:
\begin{equation}
r_\tau(i) := \E\left[\min \left\{\frac{R(i)}\tau, 1\right\}\right], \quad \forall i\in [n], \quad \mbox{ for any threshold }\tau\ge 0. \label{eq:reward-trunc}
\end{equation} 
 In order to keep the number of thresholds small (logarithmic instead of linear), we will only consider the following threshold values, also called
\emph{scales}.
$${\cal G} := 
\left\{  2^\ell \,\,:\,\, 0\le \ell \le 1+ \log_{2} W \right\}.$$ 
We still need to identify  which  scale $\tau$ to use as the threshold in \eqref{eq:reward-trunc}. To this end, 
we first classify each scale as either     {\em rich} or {\em poor}. 
Roughly speaking, in a rich scale,
the optimal knapsack solution after cost $C\cdot B$ still has large
``incremental'' reward, where $C>1$ is a parameter that will be fixed later. Formally, for any scale $\tau$, let $g_\tau$ denote the optimal LP value function when the rewards are $r_\tau$, i.e.,
$$g_\tau (D) = \textstyle \max \left\{ \sum_{i\in T} r_\tau(i)\cdot x_i \,\, \big| \,\,\sum_{i\in T} c(i)\cdot x_i \le D, \, x\in [0,1]^T\right\}.$$ 
The scale $\tau$ is called {\em rich} if the derivative $g'_\tau (C\cdot B) > \frac\epsilon{B}$, and it is called {\em poor} otherwise.  
The \emph{critical scale} is the smallest scale
$\kappa$ that is poor, and so represents a transition from rich to poor: we will show later that this always exists. Informally, we can also view the critical scale $\kappa$ as having derivative $g'_\kappa (C\cdot B) = \frac\epsilon{B}$.

For our analysis, we will choose parameter $C>1+\frac2\epsilon$.

\begin{algorithm}[h]
\caption{$\SKP(N, \mathbf{c}, \mathbf{R}, B, \epsilon)$} \label{alg:stoch-knap}
\begin{algorithmic}[1]
\State\label{step:cost} let $T = \{ i\in N \,:\, c(i) \le B\}$ 
\For{each scale $\tau \in {\cal G}$}
\State run algorithm $\A_{\rv{KS}}$ (Theorem~\ref{thm:knap}) on the deterministic knapsack instance with items $T$, rewards $r_\tau$, costs $c$ and budget $D=C\cdot B$, to obtain integral solution $S_{\tau}$  
\State if derivative $g'_\tau (C\cdot B) > \frac\epsilon{B}$ then scale $\tau$ is {\bf rich}; else it is {\bf poor}
\EndFor 
\State\label{step:critical} let {\bf critical} scale $\kappa$ be the smallest  poor  scale  in ${\cal G}$  
\State \Return items $S=S_{\kappa}$ 
\end{algorithmic}
\end{algorithm}

\medskip

We first show that the critical scale always exists. 
\begin{lemma}
The critical scale $\kappa$ in Step~\ref{step:critical} of Algorithm~\ref{alg:list} is always well defined.
\end{lemma}
\proof{Proof.}
To prove that there is a smallest poor scale, it suffices to show that not all scales can be rich. We claim that the last scale $\tau\ge W$ cannot be rich. Suppose (for a contradiction) that scale $\tau$ is rich. Then, by concavity of $g$ (see property 4 in Theorem~\ref{thm:knap}), we have $g(D)\ge D\cdot g'(D)> D\cdot \frac\epsilon{B}= C \epsilon\ge 1$. On the other hand,  the total deterministic reward at this scale, $\sum_{i=1}^n r^\tau_i \le \frac{W}{\tau} \le 1$. Thus, $g(D)\le 1$, a contradiction.  
\endproof

The next lemma proves the cost property.
\begin{lemma}\label{lem:phase-i-cost}
For any scale $\tau$, the cost $c(S_{\tau})\le (C+1)\cdot B $. Hence, the solution $S$ returned by \SKP has cost at most $(C+1)\cdot B$.   
\end{lemma}
\proof{Proof.}
Consider any scale $\tau$. We have $S_{\tau}=Q$ where $Q$ is the integer solution from Theorem~\ref{thm:knap}. It follows that $c(S_{\tau}) = c(Q) \le C B + \max_{i\in T} c_i \le (C+1)B$; note that we only consider items of cost at most $B$ (see Step~\ref{step:cost} of Algorithm~\ref{alg:stoch-knap}).  
\endproof
 
 We now prove the reward property. 
For {\em any} adaptive policy $\AD$ of cost at most $B$, w will show:
\begin{equation}
\Pr_{\AD,R}\left[ R(S) < R(\AD) \right] \le 2\epsilon, \label{eq:rew-main}
\end{equation}
where $S$ is the solution from Algorithm~\ref{alg:stoch-knap}. The probability above is taken over the randomness in the rewards $R$ as well as  the choice of items in policy $\AD$. 
This suffices to prove Theorem~\ref{thm:stoch-knap-prob}.

 Let $O\sse N$ denote the (random) items selected by adaptive policy $\AD$.   Note that $O$ is a random subset as policy $\AD$ is adaptive. On the other hand, $S\sse N$ is a deterministic subset. We need to show that the probability that $O$ has more reward than $S$ is small. 
The key idea is to use the \emph{critical scale} $\kappa$  to argue that the following hold with large enough probability: 
\begin{itemize}
    \item 
 reward of $O \setminus S$ is at most $\kappa$, and 
 \item 
 reward of $S \setminus O$ is at least $\kappa$.
\end{itemize}

 For any subset $I\sse N$  of items, we use $R(I):=\sum_{i\in I} R(i)$ to denote the total observed reward in $I$. 
We also use $\kappa^-$  to be the scale immediately preceding the critical scale $\kappa$. (If $\kappa=1$ then $\kappa^-$ is undefined, and all steps involving $\kappa^-$ can be ignored.) 

\paragraph{Upper bounding reward of $O \setminus S$.} Define new random rewards:
\begin{equation} 
    U_i =  \min\left\{\frac{R(i)}{\kappa}\,,\, 1\right\}, \quad \forall i\in N.
\end{equation}
Note that this is the truncated reward at scale $\kappa$, and  $\E[U_i]=r_\kappa(i)$ for all items $i\not\in  S$. 
 We now show that any  adaptive  policy that selects items of cost $B$ (outside $S$) cannot get too much reward.
\begin{lemma}\label{cl:opt-ub}
If ${\cal A}$ is any {\em adaptive} policy of selecting items from $N\setminus S $ with cost at most $ B$ then  $\pr_{{\cal A}, R}\left[R({\cal A}) < \kappa\right] \ge 1-\epsilon$. Hence, $\pr\left[R(O \setminus S ) \ge \kappa\right] \le \epsilon$. \end{lemma}
\proof{Proof.}
Recall the definition of subset $S$ from Algorithm~\ref{alg:stoch-knap}. This is an approximate solution to the deterministic knapsack instance on items $T$ (each of cost at most $B$) with rewards $\mathbf{r}_\kappa$, where $\kappa$ is the critical scale.  
The solution $S\sse T$ is constructed as follows (see Theorem~\ref{thm:knap}).
The items in $T$ are ordered greedily in non-increasing reward-to-cost ratio. Then,  $S$ is the minimal prefix of $T$ with total cost at least the budget $D=C B$. Moreover, the derivative $g'(D)=\frac{r_\kappa(t)}{c(t)}$, where $t$ is the last item added to $S$ (in the greedy ordering). 

As $\kappa$ is a poor scale, the derivative $g'(D)=\frac{r_\kappa(t)}{c(t)} \le \frac\epsilon{B}$. 
We now claim:
\begin{equation}
\label{eq:knap-prefix-cl}
\mbox{Any subset $A\sse N\setminus S$ with cost $c(A)\le B$ has reward }r_\kappa(A)\le \epsilon. 
\end{equation}
Clearly, every variable in $A$ has cost at most $B$: so $A\sse T\setminus S$. Using the fact that $S$ is a {\em prefix} of $T$ in  the greedy ordering (and $t$ is the last item in $S$),  
$$\frac{r_\kappa(A)}{c(A)} \le \max_{j\in A} \frac{r_\kappa(j)}{c(j)}  \le \min_{i\in S} \frac{r_\kappa(i)}{c(i)} = \frac{r_\kappa(t)}{c(t)}  \le \frac\epsilon{B}.$$
Using $c(A)\le B$, it now follows that $r_\kappa(A)\le \epsilon$, proving \eqref{eq:knap-prefix-cl}.

 Now consider the adaptive policy ${\cal A}$. We also use ${\cal A}$ to denote the (random) subset selected. Note that  ${\cal A}\sse N\setminus S$. The expected $U$-reward of this policy is:
$$ A^* = \E_{{\cal A},R}\left[\sum_{i\in {\cal A}} U_i\right] = \E_{{\cal A},R}\left[\sum_{i\in N}  \mathbf{1}_{i\in {\cal A}} \cdot U_i\right] =\sum_{i\in N}  \pr_{{\cal A}}(i\in {\cal A})\cdot \E_R[U_i] = \E_{{\cal A}}\left[\sum_{i\in {\cal A}} \E_R[U_i]\right].$$ 
The third equality above uses the fact that $\{U_i\}_{i\in N}$ are independent: so $U_i$ is independent of $i\in {\cal A}$. 
Note that every outcome $A\sse N\setminus S$ of ${\cal A}$ has total cost at most $B$. So, by \eqref{eq:knap-prefix-cl},  the total expected $U$-reward $\sum_{i\in { A}} \E_R[U_i] = r_\kappa(A)\le \epsilon$ . Combined with the above, we get 
$$A^* = \E_{{\cal A},R}\left[\sum_{i\in {\cal A}} U_i\right]\le \epsilon.$$ 

As $U_i$s are non-negative, using Markov's inequality, we have  $\pr_{{\cal A},R}\left[\sum_{i\in {\cal A}} U_i \ge 1\right]\le \epsilon$. Hence,
\begin{equation}\label{eq:opt-minus-alg}
\pr_{{\cal A},R}\left[\sum_{i\in {\cal A}} U_i < 1\right]\ge 1- \epsilon.
\end{equation}
Now, observe that $\sum_{i\in {\cal A}} U_i < 1$ implies $\sum_{i\in {\cal A}} \min(R(i),\kappa) < \kappa$, which in turn implies $\sum_{i\in {\cal A}} R(i) < \kappa$. Combined with \eqref{eq:opt-minus-alg} this proves the first part of the lemma.

For the second part of the lemma, consider   $O\setminus S $ as the adaptive policy ${\cal A}$ of cost at most $B$. It follows that 
 $\pr\left[ R(O\setminus  S ) \ge  \kappa\right] \le \epsilon$, 
as claimed. 
\endproof

\paragraph{Lower bounding reward of $S \setminus O $.} We first lower bound the {\em expected} reward.
\begin{lemma}\label{lem:s-minus-o-exp}
If $\kappa>1$ then $r_{\kappa}(S \setminus O)\ge \frac{\epsilon (C-1)}{2}$ for every outcome $O$ of $\AD$.
\end{lemma}
\proof{Proof.}
As $\kappa>1$, the rich scale $\kappa^-$ exists. To reduce notation let $r^-(i)=r_{\kappa^-}(i)$ be the reward at scale $\kappa^-$. Recall that the budget $D=C B$. Also, let $h(D)$ and $g(D)$ denote the optimal LP values of the  knapsack instances (with items $T$) in scales $\kappa^-$ and $\kappa$ respectively. 
For any item $i$, we have $r_\kappa(i) \ge \frac{\kappa^-}{\kappa}\cdot r^-(i) = \frac12\cdot  r^-(i)$.
We number  the items in $T$ in the greedy order at scale $\kappa^-$, i.e., in decreasing order of $\frac{r^-(i)}{c(i)}$. Let $t$ index the item so that the derivative $h'(D)=\frac{r^-(t)}{c(t)}=:\theta$. By Theorem~\ref{thm:knap}, the cost of the first $t$ items, $\sum_{i=1}^t c(i)\ge D$. Moreover, 
\begin{equation}\label{eq:deriv}
    \frac{r_\kappa(i)}{c(i)} \ge \frac12 \cdot\frac{r^-(i)}{c(i)}\ge \frac12 \cdot\frac{r^-(t)}{c(t)} = \frac{\theta}{2} \quad\mbox{ for all }1\le i\le t.\end{equation}
Let $\pi= \langle \pi_1, \pi_2,\cdots  ,\pi_{|T|}\rangle$ denote the items in decreasing order of the  ratio $\frac{r_\kappa(i)}{c(i)}$, which corresponds to scale $\kappa$. Note that the derivative  $g'(D)$ equals the ratio of the first item $\pi_\ell$ (in the order $\pi$) such that $\sum_{i=1}^\ell c(\pi_i) \ge D$; see Theorem~\ref{thm:knap}. From \eqref{eq:deriv} and the fact that $\sum_{i=1}^t c(i)\ge D$, it follows that  the total cost of items with ratio $\frac{r_\kappa(i)}{c(i)}$ at least $\frac{\theta}{2}$ is at least $D$. Hence, we have  $g'(D) \ge \frac{\theta}{2} = \frac{h'(D)}{2}$.

Now, using  the fact that $\kappa^-$ is rich, we have $h'(D) \ge \frac{\epsilon }{B}$. Therefore, 
$g'(D) \ge \frac{\epsilon }{2 B}$. Furthermore, using the concavity of $g$ (see Theorem~\ref{thm:knap}) and $D=C B$, 
\begin{equation}\label{eq:g-concave-rich}
g(D) \ge  g(B) + g'(D)\cdot (D - B) \ge  g(B) + \frac{\epsilon }{2 B} (D-B) = g(B)+\frac{(C-1)\epsilon}{2}.
\end{equation}
By Theorem~\ref{thm:knap} (for the knapsack instance at scale $\kappa$), the items $S$ have reward $r_\kappa(S)\ge  g(D)$. 

Now consider any outcome $O$ of the adaptive policy $\AD$. The total cost $c(O)\le B$. So, $O$ is always a feasible solution to the knapsack instance  with budget $B$. This implies $r_\kappa(O)\le g(B)$; recall that $g(B)$ is the optimal LP value. Therefore,
\begin{equation*} 
r_\kappa(  S \setminus O) \ge r_\kappa(S  ) - r_\kappa(O) \ge g(D) - g(B) \ge \frac{(C-1)\epsilon}{2}. 
\end{equation*}
The last inequality uses \eqref{eq:g-concave-rich}.
\endproof

The following is a Chernoff-type bound. 
\begin{lemma}\label{cl:alg-lb}
We have $\pr\left[R(S \setminus O) < \kappa \right]  \le e^{-({\mu} -\ln {\mu} -1)}$, where $\mu= (C-1)\epsilon/2$.  
\end{lemma}
\proof{Proof.}Recall that $U_i = \min\left\{\frac{R(i)}{\kappa} , 1\right\}$. Let $Z:=S\setminus O$. 
By Lemma~\ref{lem:s-minus-o-exp},  $\sum_{i\in Z} \E[U_i] = r_\kappa(Z) \ge \mu$. Moreover, by our choice $C>1+\frac2\epsilon$, we have $\mu>1$.   Note that $R(Z) < \kappa$ implies $\sum_{i\in Z} U_i <1$.  So it suffices to  upper bound $\pr\left[\sum_{i\in Z} U_i <1\right]$. 

Let $y>0$ be some parameter. We have:
$$\pr\left[\sum_{i\in Z} U_i < 1\right]  = \pr\left[ e^{-y\sum_{i\in Z} U_i} > e^{-y}\right] \le \E\left[ e^{-y\sum_{i\in Z} U_i}  \right]\cdot e^{y} = e^y \prod_{i\in Z} \E\left[ e^{-y U_i}  \right].$$
By convexity of $g(u) = e^{-yu}$ we have $e^{-yu} \le 1-(1-e^{-y})\cdot u$ for all $u\in [0,1]$. Taking expectation over $U_i\in [0,1]$, it follows that $\E[e^{-y U_i}] \le  1-(1-e^{-y})\cdot \E[U_i]\le \exp\left(-(1-e^{-y})\cdot \E[U_i]\right)$. 
Combined with the above,
$$\pr\left[\sum_{i\in Z} U_i <1\right]  \le e^y \prod_{i\in Z} e^{-(1-e^{-y})\cdot \E[U_i]} = e^{y-(1-e^{-y})\cdot {\mu}}.$$
Setting $y=\ln {\mu} >0$, the right-hand-side above is $e^{-{\mu}+ 1+\ln {\mu}}$ which completes the proof.
\endproof

\paragraph{Wrapping up.} 
We are now ready to finish the proof of \eqref{eq:rew-main}.  By  Lemma~\ref{cl:opt-ub}, 
$\pr\left[ R(O\setminus  S ) \ge  \kappa\right] \le \epsilon$. By Lemma~\ref{cl:alg-lb}, 
$$\pr\left[R(S \setminus O) < \kappa \right]  \le e^{-({\mu} -\ln {\mu} -1)} \le \epsilon,$$
where we used the choice $C=\Omega\left(\frac{1}{\epsilon}\ln\frac1\epsilon\right)$, which implies $\mu=\Omega\left(\ln\frac1\epsilon\right)$.
Equation~\eqref{eq:rew-main} now follows by  union bound.

\section{Score Classification Algorithm}

 \label{sec:ssclass-alg}

 We now provide a constant-factor approximation algorithm for \ssc, and prove  Theorem~\ref{thm:main}. Recall that there are two random rewards $R_0(i) = w_i(1-X_i)$
  and $R_1(i)= w_i X_i$ associated with each item $i\in [n]$. Moreover, a policy for \ssc  can make progress by collecting either $R_0$ or $R_1$ reward. So, our algorithm (described formally in Algorithm~\ref{alg:list}) divides the cost incurred equally between ``covering'' $R_0$ and $R_1$ rewards. 
 The non-adaptive list $L$ for \ssc is built  in phases. In each phase $\ci \geq 0$, we select  items of total cost $O(2^\ci)$. This is done  by invoking the stochastic knapsack algorithm (Theorem~\ref{thm:stoch-knap-prob}) twice: using $R_0$ and $R_1$ rewards separately (with  parameter $\epsilon=0.15$). Finally, 
the non-adaptive policy probes items in the order  $L$ until it identifies the class.

\begin{algorithm}[h]
\caption{Non-adaptive  algorithm for \ssc} \label{alg:list}
\begin{algorithmic}[1]
\State list $L\gets \emptyset$ 
\For{phase $\ci = 0, 1, \ldots$}
\State let $N =  [n] \setminus L $ be the remaining items
\For{$b=0,1$} 
\State \label{step:phase-rew} $L^\ci_b\gets \SKP( N, \mathbf{c}, \mathbf{R}_b, 2^\ci , 0.15)$ 
\EndFor
\State update $L\gets L\circ L^\ci_0 \circ L^\ci_1$
\EndFor
\State \Return list $L$
\end{algorithmic}
\end{algorithm}

Note that there are $O(\log (nc_{max}))$ phases in Algorithm~\ref{alg:list}, and each phase calls algorithm \SKP twice. As the running time of \SKP is nearly-linear, so is the runtime of our \ssc algorithm.


Let \NAC denote the  non-adaptive policy obtained in Algorithm~\ref{alg:list} . 
We now analyze the expected cost of \NAC. We denote by $\OPT$ an optimal adaptive solution for $\mathtt{SSClass}$. To analyze the algorithm, we use the following notation. 
\begin{itemize}
    \item $u_{\ci}$: probability that \NAC is not complete by end of phase $\ci$. 
    \item $u_{\ci}^*$: probability that $\OPT$ costs at least $2^{\ci}$.
\end{itemize}
We can assume by scaling that the minimum cost is $1$. So $u_0^* = 1$. When it is clear from context, we use $\OPT$ and $\NAC$ to also denote the  random  cost incurred by the respective policies. We also divide $\OPT$ into phases: phase $\ci$ corresponds to items in $\OPT$ after which its cumulative cost is between $2^{\ci-1}$ and $2^\ci$. So, the items selected by $\OPT$ in its first $\ci$ phases is the maximal prefix having cost at most $2^\ci$.
The following lemma forms the crux of the analysis.
\begin{lemma}[Key lemma]\label{lem:main}
For any phase $\ci \geq 1$, we have $ u_{\ci} \leq q\cdot u_{\ci-1} +  u_{\ci}^*$ where $q\le 0.3$.
\end{lemma}

Using the cost property in Theorem~\ref{thm:stoch-knap-prob}, we get:
\begin{lemma}\label{lem:nacl-cost}
The total cost of \NAC in any phase $\ci$ is at most $C\cdot 2^\ci$, where $C=O(1)$.
\end{lemma}
\proof{Proof.}
The items in phase $\ci$ of \NAC are $L^\ci_0\cup L^\ci_1$. Using Theorem~\ref{thm:stoch-knap-prob} (with $\epsilon=0.15$ and $B=2^\ci$), the total cost of $L^\ci_0$ (resp.  $L^\ci_1$) is  $O(2^\ci)$. The lemma now follows. 
\endproof 
 
We first complete the proof of Theorem~\ref{thm:main} using Lemma~\ref{lem:main}. 
\proof{Proof of Theorem~\ref{thm:main}} This proof is  fairly standard, see e.g., \cite{EneNS17}. 
By Lemma~\ref{lem:nacl-cost}, the total cost until end of \emph{phase} $\ci$ (in \NAC) is at most $ C\sum_{j=0}^\ci 2^{j} \le 2 C \cdot 2^\ci$. Let $\gamma=2 C$ below.  Moreover, \NAC ends in phase $\ci$ with probability $(u_{\ci-1} - u_{\ci})$. As a consequence of this, we have
\begin{align}
    \E[\NAC] \ &\leq \  \gamma\cdot (1 - u_{0}) +  \sum_{\ci \geq 1} \gamma\cdot 2^{\ci} \cdot (u_{\ci-1} - u_{\ci}) 
                      = \ \gamma+ \gamma\cdot \sum_{\ci \geq 0} 2^\ci u_{\ci}. \label{eq:alg-ub}
\end{align}
Similarly, we can bound the cost of the optimal adaptive algorithm as
\begin{equation}\label{eq:opt-lb}
    \E[{\OPT}] \geq \sum_{\ci \geq 0} 2^{\ci} (u_\ci^* - u_{\ci+1}^*) \geq u_0^* + \frac{1}{2}\cdot \sum_{\ci \geq 1} 2^\ci u_\ci^* = 1 + \frac{1}{2}\cdot \sum_{\ci \geq 1} 2^\ci u_\ci^*,
\end{equation}
where the final equality uses the fact that $u_{0}^* = 1$. Define $\Gamma := \sum_{\ci \geq 0} 2^\ci u_{\ci}$. We have
\begin{align*}
    \Gamma &= \sum_{\ci \geq 0} 2^\ci u_{\ci}
    \,\,\leq\,\, u_{0} + q\cdot  \sum_{\ci \geq 1} 2^\ci \cdot u_{\ci-1} +  \sum_{\ci \geq 1}2^\ci \cdot u_\ci^* \,\,\leq \,\, u_{0} +  q\cdot\sum_{\ci \geq 1} 2^\ci u_{\ci-1} + 2 \cdot (    \E[{\OPT}]  - 1)\\
    &= u_{0} + 2q\cdot \Big(\sum_{\ci\geq 0}2^\ci u_{\ci}\Big) + 2 \cdot (\E[{\OPT}]  - 1)\,\, \leq\,\, 2q\cdot \Gamma + 2\, \E[{\OPT}] - 1,
  \end{align*}
where the first inequality follows from Lemma~\ref{lem:main}, the second inequality from \eqref{eq:opt-lb}, and the last inequality from the fact that $u_{0} \leq 1$. Thus, $\Gamma \leq \frac{2}{1-2q} \cdot \E[{\OPT}] -1$. From
\eqref{eq:alg-ub}, we conclude $\E[\NAC] \leq \frac{2\gamma }{1-2q} \cdot \E[{\OPT}]$. Setting $C = O(1)$ and $q= 0.3$  completes the proof. 
\endproof

\subsection{Proof of Lemma~\ref{lem:main}}
Fix any phase $\ci \geq 1$, and let $N\sse [n]$ be the set of remaining items (i.e., items that have not been added to the list $L$ in previous phases). Let  $\sigma$ denote the realizations of the items $[n]\setminus N$, i.e., the items probed in the first $\ci-1$ phases of \NAC. We further define the following {\em conditioned on $\sigma$}:
\begin{itemize}
    \item $u_{\ci}(\sigma)$: probability that \NAC is not complete by end of phase $\ci$. 
    \item $u_{\ci}^*(\sigma)$: probability that $\OPT$ costs  at least $2^{\ci}$, i.e., $\OPT$ is not complete by end of phase $\ci$.
\end{itemize}
Note that $u_{\ci-1}(\sigma)$ is either $0$ or $1$ (as $\sigma$ contains all realizations in the first $\ci-1$ phases). If $u_{\ci-1}(\sigma)=0$ (i.e., \NAC is complete before phase $\ci$)  then $u_{\ci}(\sigma)=0$ as well: so, $    u_\ci(\sigma) \leq u_\ci^*(\sigma) + 0.3$. 
On the other hand, if  $u_{\ci-1}(\sigma) = 1$ (i.e.,  \NAC does not complete before phase $\ci$) then  we will prove
\begin{equation}\label{eq:main-lemma}
    u_\ci(\sigma) \leq u_\ci^*(\sigma) + 0.3.
\end{equation}
This would imply  $u_\ci(\sigma) \leq u_\ci^*(\sigma) + 0.3 u_{\ci-1}(\sigma)$ for {\em all} realizations $\sigma$. 
Taking expectation over $\sigma$ then proves Lemma~\ref{lem:main}. It
remains  to prove Equation \eqref{eq:main-lemma}. 

We denote by $\cR_0$ and $\cR_0^*$ the total $R_0$ reward obtained in the
first $\ci$ phases by \NAC and $\OPT$ respectively. We similarly define
$\cR_1$ and $\cR_1^*$. To prove Equation \eqref{eq:main-lemma}, we
first use the reward-property in Theorem~\ref{thm:stoch-knap-prob} to show that the probabilities 
$\pr(\cR_0^* > \cR_0)$ and $\pr(\cR_1^* > \cR_1)$ are
small. 
\begin{lemma}\label{lem:rew-prob}
For $b\in \{0,1\}$, we have $ \pr(\cR_b < \cR_b^* \mid \sigma) \leq 0.15$.
\end{lemma}
\proof{Proof.} Fix any $b\in \{0,1\}$ and condition on realization $\sigma$. Let $R_b(\sigma)$ denote the total $R_b$ reward from the realizations in $\sigma$; note that this is a deterministic value as we have conditioned on $\sigma$. Let $S=L^\ci_b\sse N$ be the items added in phase $\ci$ corresponding to $R_b$-rewards (see Step~\ref{step:phase-rew} in Algorithm~\ref{alg:list}). 

Observe that  $\cR_b\ge  R_b(\sigma) + R_b(S)$ as $\sigma$ corresponds to all items in the first $\ci-1$ phases and the items selected in phase $\ci$ is a superset of $S$. (Recall that all rewards are non-negative.) 

Let $\AD$ denote the adaptive policy obtained by  restricting $\OPT$ (conditional on $\sigma$) to its first $\ci$ phases and items $N$. 
 In other words, if $\OPT$ selects any item $i\in [n]\setminus N$ (in its first  $\ci$ phases) then $\AD$ does not collect the reward of item $i$ but it continues to follow $\OPT$ according to $i$'s realization in $\sigma$. Note that $R_b(\AD)\ge \cR^*_b - R_b(\sigma)$. Moreover,  
  the cost of $\AD$ is at most $2^\ci$. 
  
Using the fact that $S$ is the solution of \SKP with items $N$, rewards $R_b$, budget $B=2^\ci$ and $\epsilon=0.15$ (and Theorem~\ref{thm:stoch-knap-prob}), we have $\Pr[R_b(S) < R_b(\AD)] \le 0.15$. 

Combined with $\cR_b\ge  R_b(\sigma) + R_b(S)$  and $R_b(\AD)\ge \cR^*_b - R_b(\sigma)$, we have (conditioned on $\sigma$), 
$$\Pr[\cR_b  < \cR_b^*] \le \Pr[R_b(\sigma) + R_b(S) < R_b(\sigma)+ R_b(\AD)] = \Pr[R_b(S) < R_b(\AD)] \le 0.15.$$
\endproof

Using this  lemma, we prove Equation~\eqref{eq:main-lemma}. 
\proof{Proof of Equation \eqref{eq:main-lemma}.}
Recall that we associate a pair $(\beta_j^0, \beta_j^1)$ with every class $j$. If $\OPT$ finishes in phase $\ci$, then there exists some $j$ such that $\cR_0^* \geq \beta^0_j$ and $\cR_1^* \geq \beta^1_j$. 
Thus, $$\pr(\OPT \text{ finishes by phase } \ci \mid \sigma) = 1 - u_\ci^*(\sigma) = \pr(\exists j : \cR_0^* \geq \beta_j^0 \text{ and } \cR_1^* \geq \beta_j^1 \mid \sigma).$$ 

From Lemma~\ref{lem:rew-prob} and union bound, we have $\pr(\cR_0 < \cR_0^* \text{ or } \cR_1 < \cR_1^*\mid \sigma) \leq 0.3$. Then, we have 
\begin{align*}
1 - u_\ci(\sigma)  & = \pr(\NAC \text{ finishes by phase } \ci \mid \sigma) \\&\ge \pr\left((\OPT \text{ finishes by phase } \ci) \bigwedge \cR_0 \ge \cR_0^*  \bigwedge \cR_1 \ge \cR_1^* \, \big| \,\sigma\right) \\
&\ge \pr(\OPT \text{ finishes by phase } \ci \mid \sigma) - \pr(\cR_0 < \cR_0^* \text{ or } \cR_1 < \cR_1^*\mid \sigma) \\
&\ge (1 - u_\ci^*(\sigma)) - 0.3
\end{align*}
Upon rearranging, this gives $u_\ci(\sigma) \leq  u^*_\ci(\sigma) + 0.3$ as desired.  
\endproof

\section{Explainable Stochastic $d$-Halfspace Evaluation}\label{sec:general}

We now consider \exsfe and prove  \Cref{thm:exsfe-main}. Recall that an instance of \exsfe\ also involves $n$ random items $\{X_i\}_{i=1}^n$, but the function to be evaluated is different. In particular, there are $d$ halfspaces $h_k(X) = \langle \sum_{i=1}^n w_{ki}X_i \ge \alpha_k\rangle$ for $k\in [d]$. There is also an  aggregation
    function  $f: \{0,1\}^d \to \{0,1\}$, and we want to evaluate
  $f(h_1(X), \ldots, h_d(X))$.   
  
Furthermore, the problem asks for an ``explanation'', or a witness of
  this evaluation. To define a witness, consider a tuple
  $(S, v, T)$, where $S \sse [n]$ is the set of probed items,
  $v = \langle v_i : i\in S\rangle$ are the realizations of these items, and $T \sse [d]$
  is a subset of halfspaces. This tuple $(S, v , T)$ is a
  \emph{witness} for $f\circ h$ if the following conditions are satisfied:
  \begin{itemize}
      \item   The  realizations $X_i=v_i$ of the probed items $i\in S$   determine $h_k(X)$ for all halfspaces $k\in T$. In other words, we can evaluate all the halfspaces $T\sse [d]$ using {\em only} the realizations of $S$. 
      \item The values of $\{h_k(X)\}_{k \in T}$  completely determine $f \circ h(X)$. That is, $f(y)$ has the same value for {\em every} $y\in \{0,1\}^d$ with $y_k=h_k(X)$ for $k\in T$.
        \end{itemize}
In other words, a witness is a proof of the function value that {\em only} makes use of probed variables.  
 The goal is to evaluate $f\circ h$ along with a witness (as defined above), at the minimum expected
  cost.   We assume that checking 
   whether a given tuple 
   is a witness 
   can be done efficiently.

The goal is to design a probing strategy of minimum expected cost that determines $f\circ h$ along with a witness.  
Before  describing our algorithm, we highlight the role of a witness in stochastic $d$-halfspace evaluation (by comparing to a model without witnesses). We also discuss the complexity of verifying witnesses for $f \circ h$.

{\bf Solutions with/without witness.} Solutions that are not required to provide a witness for their evaluation stop when they can infer that the  function  $f \circ h$ remains constant irrespective of the realizations of the remaining variables. For example, consider the stochastic intersection of halfspaces problem (as studied in \cite{BlancLT21}). A feasible solution without a witness can stop probing when it determines that, with probability one,  either halfspace  $h_{1}$ or halfspace $h_{2}$ is violated   (though it does not know precisely which halfspace is violated). Such a ``stopping rule'' may not be useful in situations where one also wants to know the identity of a violated halfspace (say, in order to take some corrective action). A solution with a witness (as required in our model) would provide one specific violated halfspace or conclude that all halfspaces are satisfied. 


{\textbf{Verifying witnesses.}} We now address the issue of verifying whether a tuple $(S, v , T)$ is a witness for $f \circ h$. Note that it is easy to check whether the values $X_i=v_i$ for $i\in S$ suffice to evaluate all the halfspaces in $T$. The challenge in verifying witnesses lies in confirming whether the values of $\{h_k(X)\}_{k \in T}$ completely determine $(f \circ h)(X)$. For some functions $f$, such as intersection (i.e., all halfspaces must be satisfied) or $p$-of-$d$ functions (i.e., at least $p$ of the $d$ halfspaces must be satisfied), this can be done efficiently. However, for a general aggregation function $f$, verifying a witness may require the evaluation of $f$ at all $2^{d}$ points. 
While our algorithm works for any aggregation function $f$, for a  polynomial running time, we need to assume   an efficient oracle  for verifying witnesses.  

\ignore{In this case, we can use random sampling to verify witnesses within an $\epsilon$ probability of error for any $\epsilon > 0$. Thus, we obtain the following result.

\begin{theorem}\label{thm:exsfe-oracle}
Given any $\delta > 0$, there is an algorithm for \exsfe which runs in time $\text{poly}\left(\frac{1}{\delta}, n\right)$ and provides a correct solution with cost at most $O(d^2 \log d)$ times the cost of an optimal adaptive solution with probability at least $1 - \delta$. 
\end{theorem}

We provide details in \S\ref{app:error-oracle}.} 

Our algorithm for  \exsfe  is similar to that for \ssc, and is described in Algorithm~\ref{alg:ex-halfspace}. We introduce $2d$ random rewards that help in evaluating the $d$ different  halfspaces: 
$$R_{k, 0}(i) := w_{ki}(1 - X_i) \quad and \quad R_{k, 1}(i) := w_{ki} X_i, \qquad \forall k\in [d].$$ 
Then, we ensure that the cost incurred is spread uniformly to cover these $2d$ different rewards. Again, the list-building algorithm  
proceeds in phases, where in each phase $\ell$, we utilize the stochastic knapsack algorithm (Theorem~\ref{thm:stoch-knap-prob}) with budget $2^\ell$ for each of the $2d$ rewards. This time, we set the parameter $\epsilon=\frac{0.15}d$ which enables us to show that our algorithm makes progress in covering {\em all} the $2d$ rewards. Finally, the non-adaptive policy just probes items in the order of list $L$ until the realizations of the probed items form a witness for $f \circ h$ (which is verfied using the oracle). 

\begin{algorithm}[h]
\caption{Non-adaptive algorithm for \exsfe} \label{alg:ex-halfspace}
\begin{algorithmic}[1]
\State list $L\gets \emptyset$ 
\For{phase $\ci = 0, 1, \ldots$}
\State let $N =  [n] \setminus L $ be the remaining items
\For{$k\in [d]$ and $b\in \{0,1\}$} 
\State \label{step:phase-rew-2} $L^\ci_{k,b} \gets \SKP( N, \mathbf{c}, \mathbf{R}_{k,b}, 2^\ci , \frac{0.15}d)$ 
\State update $L\gets L\circ L^\ci_{k,b}$
\EndFor
\EndFor
\State \Return list $L$
\end{algorithmic}
\end{algorithm}

The analysis is also similar to that for \ssc. Let \NAC denote the non-adaptive policy in Algorithm~\ref{alg:ex-halfspace} and $\OPT$   the optimal adaptive policy for \exsfe. We also define:
\begin{itemize}
    \item $u_{\ci}$: probability that \NAC is not complete by end of phase $\ci$. 
    \item $u_{\ci}^*$: probability that $\OPT$ costs at least $2^{\ci}$.
\end{itemize}
Again, we assume by scaling that the minimum cost is $1$; so $u_0^* = 1$. We use $\OPT$ and $\NAC$ to also denote the  random  cost incurred by the respective policies. We also divide $\OPT$ into phases: phase $\ci$ corresponds to items in $\OPT$ after which its cumulative cost is between $2^{\ci-1}$ and $2^\ci$. As before, we will show: 
\begin{lemma}\label{lem:main-2}
For any phase $\ci \geq 1$, we have $ u_{\ci} \leq q\cdot u_{\ci-1} +  u_{\ci}^*$ where $q\le 0.3$.
\end{lemma}

Using the cost property in Theorem~\ref{thm:stoch-knap-prob}, we get:
\begin{lemma}\label{lem:nacl-cost-2}
The total cost of \NAC in any phase $\ci$ is at most $C\cdot 2^\ci$, where $C=O(d^2\ln d)$.
\end{lemma}
\proof{Proof.}
The items in phase $\ci$ of \NAC are $\bigcup_{k=1}^d \left(L^\ci_{k,0}\cup L^\ci_{k,1}\right)$. For each $k\in [d]$ and $b=0,1$, using Theorem~\ref{thm:stoch-knap-prob} (with $\epsilon=\frac{0.15}d$ and $B=2^\ci$), we obtain that the cost of subset $L^\ci_{k,b}$  is  $O(\frac1\epsilon \ln \frac1\epsilon)\cdot B=O(d\ln d)\cdot 2^\ci$. The lemma now follows by combining the $2d$ subsets $L^\ci_{k,b}$. 
\endproof

The proof of Theorem~\ref{thm:exsfe-main} uses Lemma~\ref{lem:main-2} and is identical to the proof of Theorem~\ref{thm:main} (in \S\ref{sec:ssclass-alg}). The only difference is that we have $C=O(d^2\ln d)$ from Lemma~\ref{lem:main-2}, which results in the $O(d^2\ln d)$ approximation guarantee.  

It now remains to prove Lemma~\ref{lem:main-2}.  
Fix any phase $\ci \geq 1$, and let $N\sse [n]$ be the set of remaining items. Let  $\sigma$ denote the realizations of the items $[n]\setminus N$. As before, we also define the following {\em conditioned on $\sigma$}:
\begin{itemize}
    \item $u_{\ci}(\sigma)$: probability that \NAC is not complete by end of phase $\ci$. 
    \item $u_{\ci}^*(\sigma)$: probability that $\OPT$ costs  at least $2^{\ci}$, i.e., $\OPT$ is not complete by end of phase $\ci$.
\end{itemize}
As in the proof of Lemma~\ref{lem:main}, it suffices to prove
\begin{equation}\label{eq:main-lemma-2}
    u_\ci(\sigma) \leq u_\ci^*(\sigma) + 0.3, \qquad \mbox{for any $\sigma$ such that \NAC doesn't complete before phase }\ell.
\end{equation}

To this end, for any $k\in [d]$ and $b\in \{0,1\}$, let $\cR_{k,b}$ and $\cR_{k,b}^*$ be the total $R_{k,b}$ reward obtained in the
first $\ci$ phases by \NAC and $\OPT$ respectively. We
first use the reward-property in Theorem~\ref{thm:stoch-knap-prob} to show: 
\begin{lemma}\label{lem:rew-prob-2}
For any $k\in [d]$ and $b\in \{0,1\}$, we have $ \pr(\cR_{k,b} < \cR_{k,b}^* \mid \sigma) \leq \frac{0.15}d$.
\end{lemma}
\proof{Proof.} This is identical to the proof of Lemma~\ref{lem:rew-prob}, where we use random rewards $R_{k,b}$ (instead of $R_b$) and value $\epsilon=\frac{0.15}d$ (instead of $0.15$).  
\endproof

Using this  lemma, we prove Equation~\eqref{eq:main-lemma-2}.  
If $\OPT$ finishes in phase ${\ell}$, then there exists some witness $(S^{\star}, v^{\star}, T^{\star})$ where (i) $S^{\star}$ is the set of items probed by $\OPT$ until phase $\ell$,  (ii) the realizations $v^{\star}$ of the items $S^{\star}$ determine $h_k(X)$ for all halfspaces $k \in T^{\star}$, and (iii) the values $\{h_k(X)\}_{k \in T^{\star}}$ completely determine $f \circ h$. We further partition $T^\star$ into 
$$T^\star_0=\{ k\in T^\star : h_k(X)=0\} \,\, and \,\, T^\star_1=\{ k\in T^\star : h_k(X)=1\}.$$ 

For any halfspace $k\in[d]$, define thresholds:
$$\beta_{k,0} = \sum_{i=1}^n w_{ki} -\alpha_k +1 \quad and \quad \beta_{k,1}=\alpha_k.$$
Note that for $b\in \{0,1\}$, we have $h_k(X)=b$ if and only if $R_{k,b}(X) \ge \beta_{k,b}$. 

Now, using the definition of $T^\star_0$ and $T^\star_1$, we have:
\begin{enumerate}
    \item[C1.] $\cR_{k, 0}^*\ge \beta_{k,0}$  for each $k\in T^\star_0$,  and
    \item[C2.]  $\cR_{k, 1}^*\ge \beta_{k,1}$  for each $k\in T^\star_1$.  
\end{enumerate}
Let ${\cal T}$ denote the set of 3-way-partitions $(T_0, T_1, [d]\setminus T_0\setminus T_1)$ of the $d$ halfspaces such that $f$ is completely determined by setting the coordinates in $T_0$ to $0$ and those in  $T_1$ to $1$. Note that for any witness as above, we have $(T^\star_0,T^\star_1,[d]\setminus T^\star_0\setminus T^\star_1) \in {\cal T}$. 
Thus, 
\begin{align*}
1 - u_{\ell}^*(\sigma)   \, &=\, \pr\left[\OPT \text{ finishes by phase } {\ell} \,\mid\, \sigma\right] \\
		&= \pr\left[\exists (T^\star_0,T^\star_1,[d]\setminus T^\star_0\setminus T^\star_1) \in {\cal T} \,:\, \text{ conditions C1 and C2 hold }  \mid \, \sigma\right].
\end{align*}
Note that if $\OPT$  finishes by phase  ${\ell}$ and $\cR_{k,b} \ge \cR_{k,b}^*$ for all $k\in [d]$ and $b\in \{0,1\}$, then we can conclude that \NAC also finishes by  phase  ${\ell}$ (with the same witness as $\OPT$). 
By Lemma~\ref{lem:rew-prob-2} and union bound, we have $\pr\left[\exists k, b: \cR_{k,b} < \cR_{k,b}^* \mid \sigma\right] \leq 0.3$. 
Hence,  
\begin{align*}
1 - u_{\ell}(\sigma)  & = \pr(\NAC \text{ finishes by phase } {\ell} \mid \sigma) \ge \pr\left((\OPT \text{ finishes by phase } {\ell}) \bigwedge \wedge_{k, b} (\cR_{k,b} \ge \cR_{k, b}^*) \big| \,\sigma\right) \\
&\ge \pr(\OPT \text{ finishes by phase } {\ell} \mid \sigma) - \pr(\exists k, b: \cR_{k,b} < \cR_{k,b}^* \mid \sigma) \\
&\ge (1 - u_{\ell}^*(\sigma)) - 0.3
\end{align*}
Upon rearranging, this gives $u_{\ell}(\sigma) \leq  u^*_{\ell}(\sigma) + 0.3$, which proves \eqref{eq:main-lemma-2}.


\section{Sequential Testing with Batched Costs}\label{sec:batched-ssclass}

In this section, we consider \ssc and \exsfe under batch-costs and prove Theorem~\ref{thm:batched-ssc-main}. 
Recall that the batched cost structure involves an additional setup cost $\rho$, where the cost to simultaneously test a subset $S\sse[n]$ is $\rho+\sum_{i\in S}c_i$. The goal is  to evaluate the function $f$ (corresponding to \ssc or \exsfe) at the minimum expected cost.  

A solution in the batched setting is an adaptive sequence of {\em subsets} $\langle S_1, S_2,\cdots \rangle$, where all items in $S_t$ are tested simultaneously in step $t$. Note that these subsets are pairwise disjoint. If the function value $f(X)$ 
is determined after step $t$ (i.e., using the outcomes of items $S_1\cup S_2\cup \cdots S_t$) then we stop after this  step;  else, we proceed to step $t+1$ by testing $S_{t+1}$. 
Our algorithm is again non-adaptive:  it fixes the sequence of subsets upfront, and the random outcomes are only used to determine when to stop. We divide the cost incurred by any solution into its {\em total setup cost} (i.e.,  $\rho$ times the number of batches used)  and {\em total testing cost} (i.e.,  sum of costs $c_i$ over all items tested).   

Our algorithm is a simple extension of Algorithms \ref{alg:list} and \ref{alg:ex-halfspace} for \ssc and \exsfe under the usual cost structure (no batches). Recall that both Algorithms \ref{alg:list} and \ref{alg:ex-halfspace} proceed in phases, where the items $L_\ci$ selected in each phase $\ci$ have cost $\sum_{i\in L_\ci} c_i \le C\cdot 2^\ci$. The multiplier $C=O(1)$ for \ssc (Algorithm \ref{alg:list}) and $C=O(d^2\log d)$ for \exsfe (Algorithm \ref{alg:ex-halfspace}). Let $\tau=\lfloor \log_2 \rho\rfloor$. Then, the batched-cost algorithm does the following. For each phase $\ci = \tau, \tau+1,\cdots$, test subset $L_\ci$ in one batch. Note that the difference from the previous algorithms is that we start directly in phase $\tau\approx \log \rho$ instead of phase $0$ (and we test all items in a phase simultaneously). The reason we start with phase $\tau$ is that we do not want to incur the setup cost before the total testing cost is $\rho$.

\paragraph{Analysis.}  Let \NAC denote the non-adaptive policy above and $\OPT$   the optimal adaptive policy, both under batch-costs. As before, we define:
\begin{itemize}
    \item $u_{\ci}$: probability that \NAC is not complete by end of phase $\ci$. 
    \item $u_{\ci}^*$: probability that $\OPT$ costs at least $2^{\ci}$.
\end{itemize}
Using Lemmas \ref{lem:main} and \ref{lem:main-2} directly, we get:
\begin{lemma}\label{lem:main-batch}
For any $\ci \geq \tau+1$, we have $ u_\ci \leq qu_{\ci-1} + u_\ci^*$ where $q \le 0.3$
\end{lemma}
By Lemmas \ref{lem:nacl-cost} and \ref{lem:nacl-cost-2}, we get:
\begin{lemma}\label{lem:nacl-cost-batch}
The total testing cost of \NAC in any phase $\ci$ is at most $C\cdot 2^\ci$.
\end{lemma}

We are now ready to prove Theorem~\ref{thm:batched-ssc-main}.
 Note that $\OPT$ uses at least one batch: so its setup cost is at least $\rho$. The testing cost can be lower bounded as before, to get:
 \begin{align}
    \E[{\OPT}] &\geq \rho \,+\, \sum_{\ci \geq 0} 2^{\ci} (u_\ci^* - u_{\ci+1}^*)    \geq \rho \,+\, u_0^* + \frac{1}{2}\sum_{\ci \geq 0} 2^{\ci} u_{\ci^*}  \geq 1 + \rho \, + \,\frac{1}{2}\sum_{\ci \geq \tau+1} 2^{\ci} u_{\ci}^* \label{eq:batched-opt-lb}
\end{align}
where we used the fact that $u_0^* = 1$. 

Observe that with probability $1-u_{\tau}$, \NAC completes in phase $\tau$ and has total cost at most $\rho + C\cdot 2^{\tau}$. Moreover, for any $\ci > \tau$, \NAC completes in phase $\ci$ with probability $u_{\ci-1} - u_{\ci}$ and has setup cost $(\ci-\tau+1)\cdot \rho$  (corresponding to the number of batches used) and  testing cost at most $2C\cdot 2^{\ci}$ (by Lemma~\ref{lem:nacl-cost-batch}). Hence, we have
\begin{align}
    \E[\NA] \ &\leq \  (\rho + C\cdot 2^{\tau}) \cdot (1 - u_{\tau}) +  \sum_{\ci \geq \tau+1} (C\cdot 2^{\ci+1} + (\ci-\tau+1)\rho) \cdot (u_{\ci-1} - u_{\ci}) \notag\\
    &= \rho + \rho \cdot \sum_{\ci \geq \tau} u_{\ci} + C 2^{\tau} +  C \cdot\sum_{\ci \geq \tau} 2^{\ci}u_{\ci}   \,\,\leq \,\,2\rho + \sum_{\ci \geq \tau+1} 2^{\ci} u_{\ci} +C \rho + C  \cdot\sum_{\ci \geq \tau} 2^{\ci}u_{\ci}, \label{eq:batched-1}
\end{align}
where the last inequality uses $\tau \leq \log_2(\rho)\le \tau+1$.

Define $\Gamma := \sum_{\ci \geq \tau} 2^\ci u_{\ci}$. We have
\begin{align*}
    \Gamma &= \sum_{\ci \geq \tau} 2^\ci u_{\ci}
    \,\,\leq\,\,  2^{\tau}\cdot u_{\tau} + q\cdot  \sum_{\ci \geq \tau+1} 2^\ci \cdot u_{\ci-1} +  \sum_{\ci \geq \tau+1}2^\ci \cdot u_\ci^* \,\,\leq\,\,  u_{\tau} +  2q\cdot  \sum_{\ci\geq \tau}2^\ci u_{\ci} + 2 \cdot (    \E[{\OPT}]  - (\rho+1))\\
    &= u_{\tau} +  2q\cdot \Gamma + 2 \cdot (    \E[{\OPT}]  - (\rho+1)) \,\,\leq\,\, 2q\cdot \Gamma + 2\, \E[{\OPT}] - 2\rho - 1,
  \end{align*}
where the first inequality follows from Lemma~\ref{lem:main-batch}, the second inequality from \eqref{eq:batched-opt-lb}, and the last inequality from the fact that $u_{\tau} \leq 1$. Thus, $\Gamma \leq \frac{1}{1-2q} \cdot \left(2\E[{\OPT}] - 2\rho - 1\right)$. From
\eqref{eq:batched-1}, we conclude $$\E[\NA] \leq (C+2)\cdot \rho + (C+1) \cdot \frac{1}{1-2q} \cdot \left(2\E[{\OPT}] - 2\rho - 1\right) \leq  \frac{2(C+1)}{1-2q} \cdot \E[{\OPT}].$$ Setting $C = O(1)$ for \ssc and $C=O(d^2\log d)$ for \exsfe, completes the proof of Theorem~\ref{thm:batched-ssc-main}.



\section{Computational Results}\label{sec:computations}
We provide a summary of computational results of our non-adaptive algorithm for the stochastic score classification problem.  We conducted all of our computational experiments using Python 3.8 with a 2.3 GHz Intel Core i5 processor and 16 GB 2133MHz LPDDR3 memory. 
We use synthetic data to generate instances of $\mathtt{SSClass}$ for our experiments.

{
\paragraph{Instance Type.} We test our algorithm for three 
types of instances:   stochastic halfspace evaluation (\she),   unweighted stochastic score classification, and (weighted) stochastic score classification (\ssc).
The cutoffs and the weights for the score function are selected based on the type of instance being generated and will be specified later.
}

\paragraph{Instance Generation.} We test our algorithm on synthetic data generated as follows. We first set $n \in \{100, 200, \ldots, 1000\}$. Given $n$, we generate $n$ Bernoulli variables, each with probability chosen  uniformly from $(0, 1)$. We set the costs of each variable to be an integer in $[10, 100]$. To select cutoffs (when $B \neq 2$), we first select $B \in \{5, 10, 15\}$ and then select the cutoffs  (based  on  the  value  of $B$)  uniformly  at  random  in  the  score  interval. 
For each $n$ we generate $10$ instances. For each instance, we sample $50$ realizations in order to calculate the  average cost and average runtime.

\paragraph{Algorithms.} We compare our non-adaptive $\mathtt{SSClass}$ algorithm (Theorem~\ref{thm:main}) against a number of prior algorithms. 
For \she instances, we compare to the {\em adaptive} 3-approximation algorithm from \cite{DeshpandeHK16}. 
For {\em unweighted}   $\mathtt{SSClass}$ instances, we compare to the non-adaptive algorithm from \cite{GkenosisGHK18}, 
which was shown to be a $6$-approximation to the optimal adaptive policy by \cite{PlankSchewior22-arxiv} and \cite{Liu22-arxiv}.
For general   $\mathtt{SSClass}$ instances, we compare to the {\em adaptive} $O(\log W)$-approximation algorithm from \cite{GkenosisGHK18}.  
{
We would like to emphasize that we compare our single algorithm with
other algorithms that are tailored to specific cases of \ssc.
}
As a benchmark, we also compare to  a naive non-adaptive algorithm that probes variables in a random order. 
We also compare to an information-theoretic lower bound (no adaptive policy can do better than this lower bound). 
We obtain this lower bound by using an integer linear program to compute the (offline) optimal probing cost for a given realization (see \S\ref{sec:comp-lb} for details), and then taking an average  over $50$ realizations.

\ignore{ 
\paragraph{Parameters $C$, $\delta$, and $\theta$.} As noted in \S\ref{sec:ssclass-alg}, our algorithm achieves a constant factor approximation guarantee for any constant $C > 1$, $\delta \in (0, 1)$ and $\theta >1$. 
For our final computations, we (arbitrarily) choose values $C=2$, $\delta=0.01$, and $\theta=2$. 
}

\paragraph{Reported quantities.} 
For every instance, we compute the cost and runtime of each algorithm  by taking an average over  $50$ independent realizations. 
For the non-adaptive algorithms, note that we only need {\em one}  probing sequence  for each instance (irrespective of the realization). On the other hand, adaptive algorithms need to find the probing sequence afresh for each realization. 
As seen in all the runtime plots, the non-adaptive algorithms are significantly faster.

\begin{figure}[h!]
     \centering
     \begin{subfigure}[b]{0.41\textwidth}
         \centering
        \includegraphics[width=2.5in]{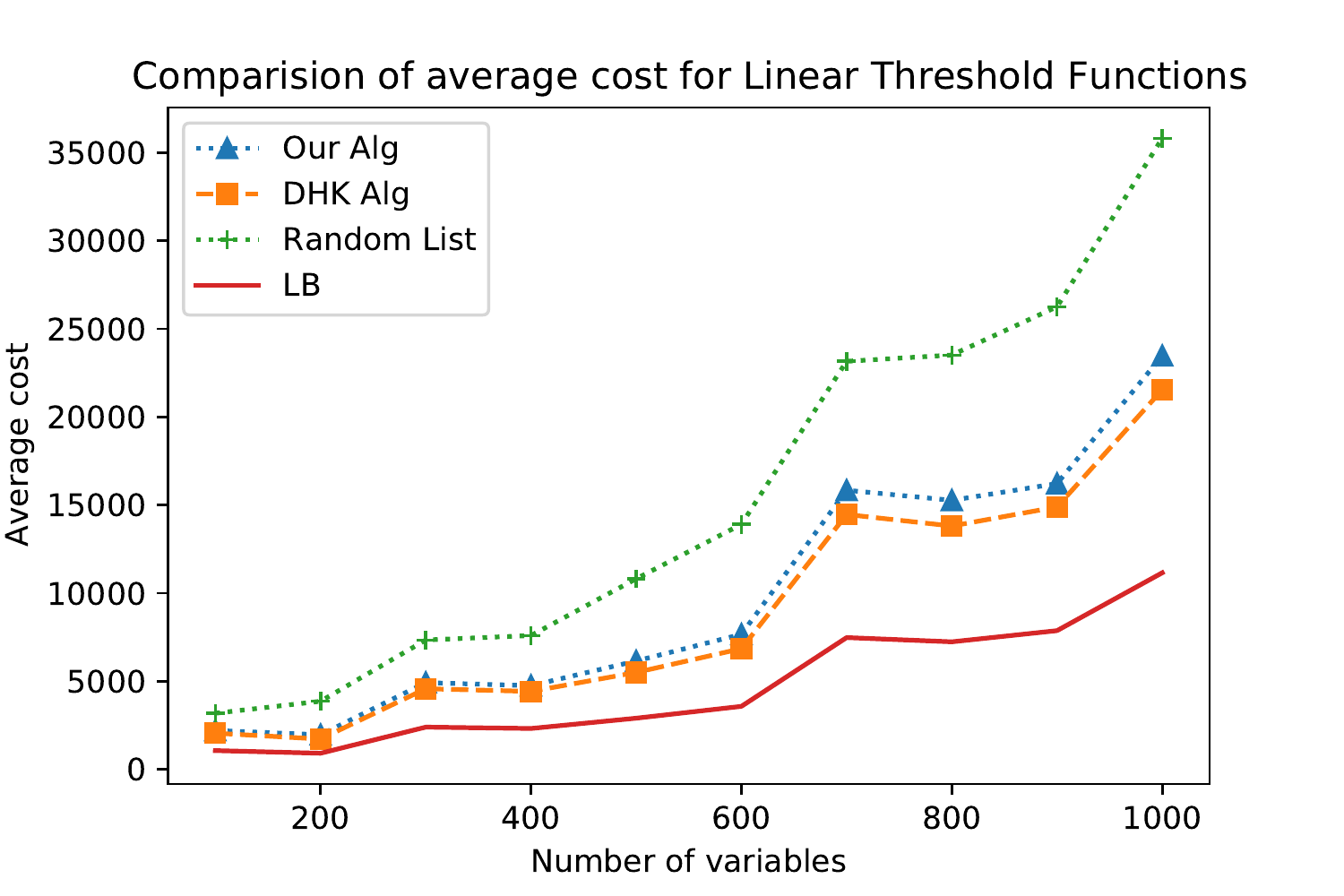}
     \end{subfigure}
     \qquad
     \begin{subfigure}[b]{0.41\textwidth}
         \centering
        \includegraphics[width=2.5in]{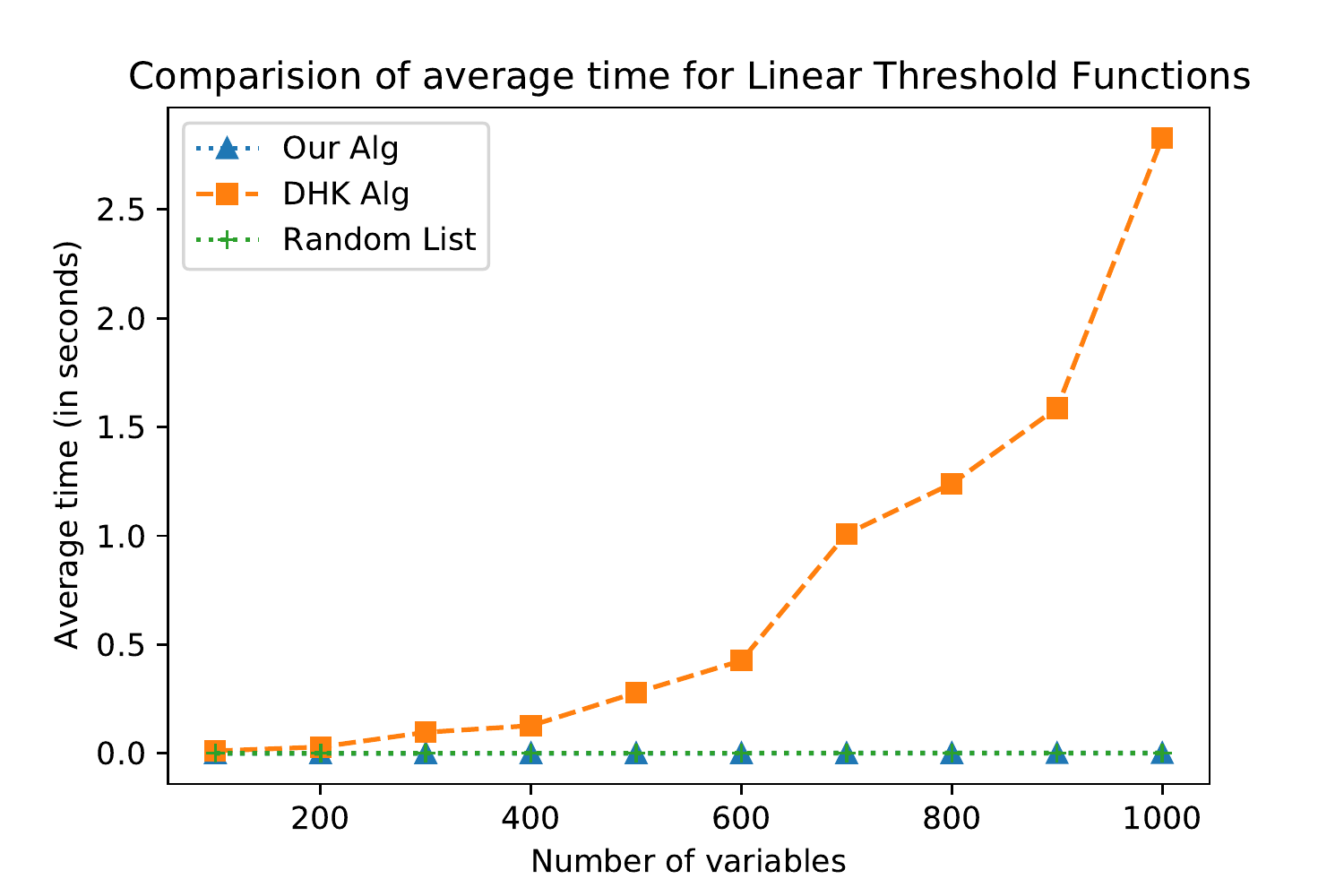}
     \end{subfigure}
     \caption{Cost and runtime results for \she}
        \label{fig:plots-syn-LTFs}
\end{figure}

\begin{figure}[h!]
     \centering
     \begin{subfigure}[b]{0.3\textwidth}
         \centering
         \includegraphics[width=2in]{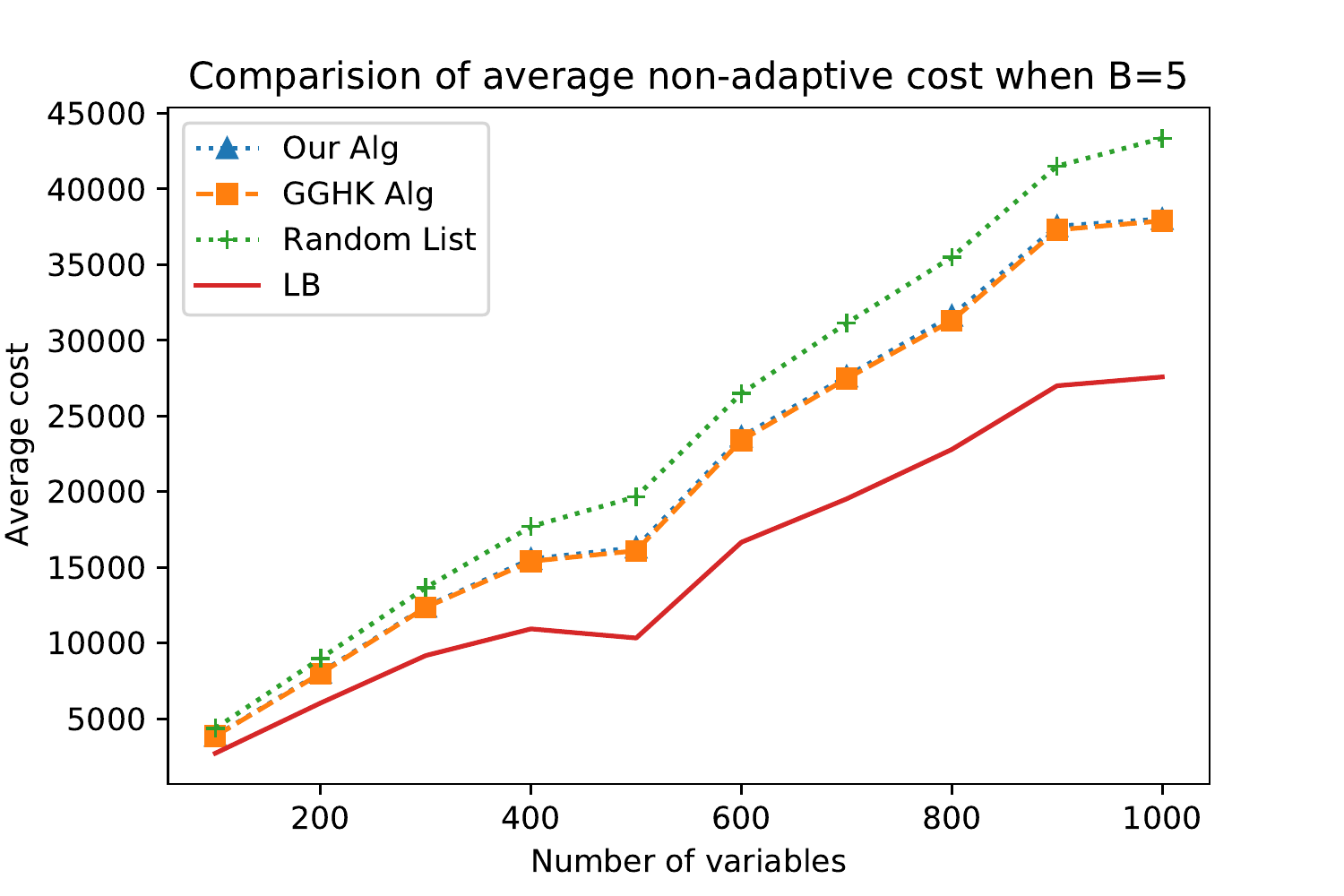}
     \end{subfigure}
     \quad
     \begin{subfigure}[b]{0.3\textwidth}
         \centering
         \includegraphics[width=2in]{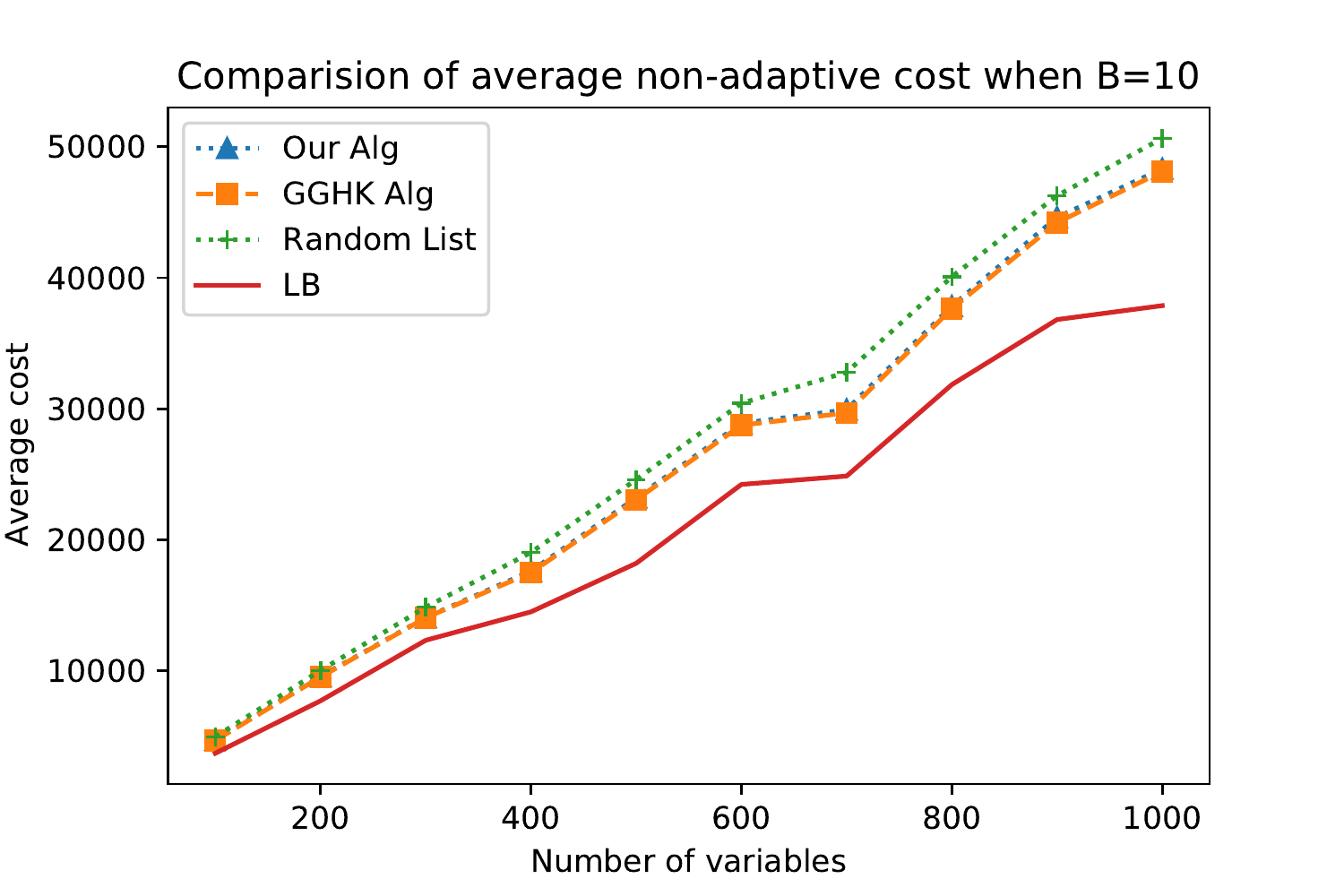}
     \end{subfigure}
     \quad
     \begin{subfigure}[b]{0.3\textwidth}
         \centering
         \includegraphics[width=2in]{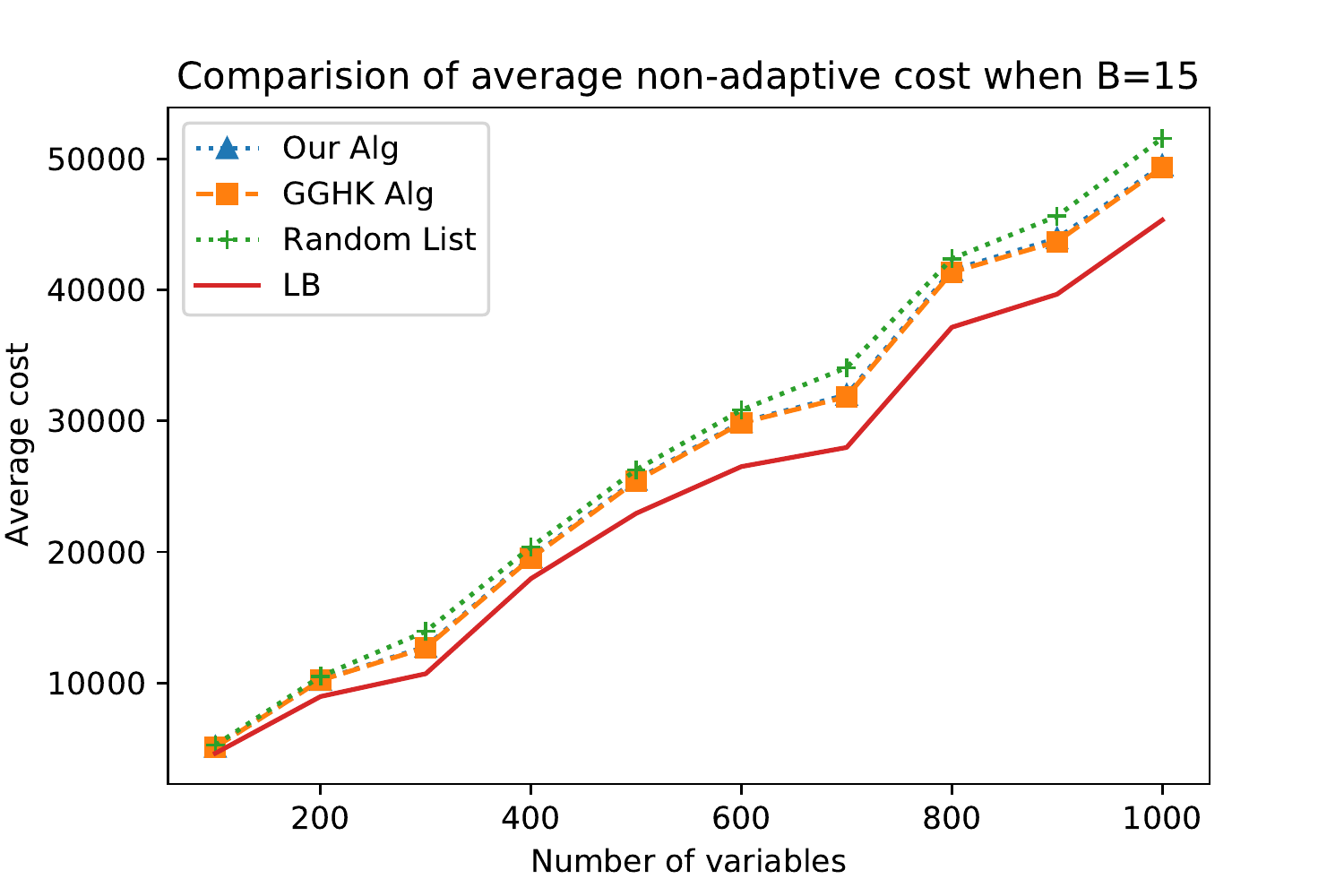}
     \end{subfigure}
     \caption{Cost results for Unweighted $\mathtt{SSClass}$ }
        \label{fig:plots-syn-unweighted}
\end{figure}

\begin{figure}[h!]
     \centering
     \begin{subfigure}[b]{0.28\textwidth}
         \centering
         \includegraphics[width=2in]{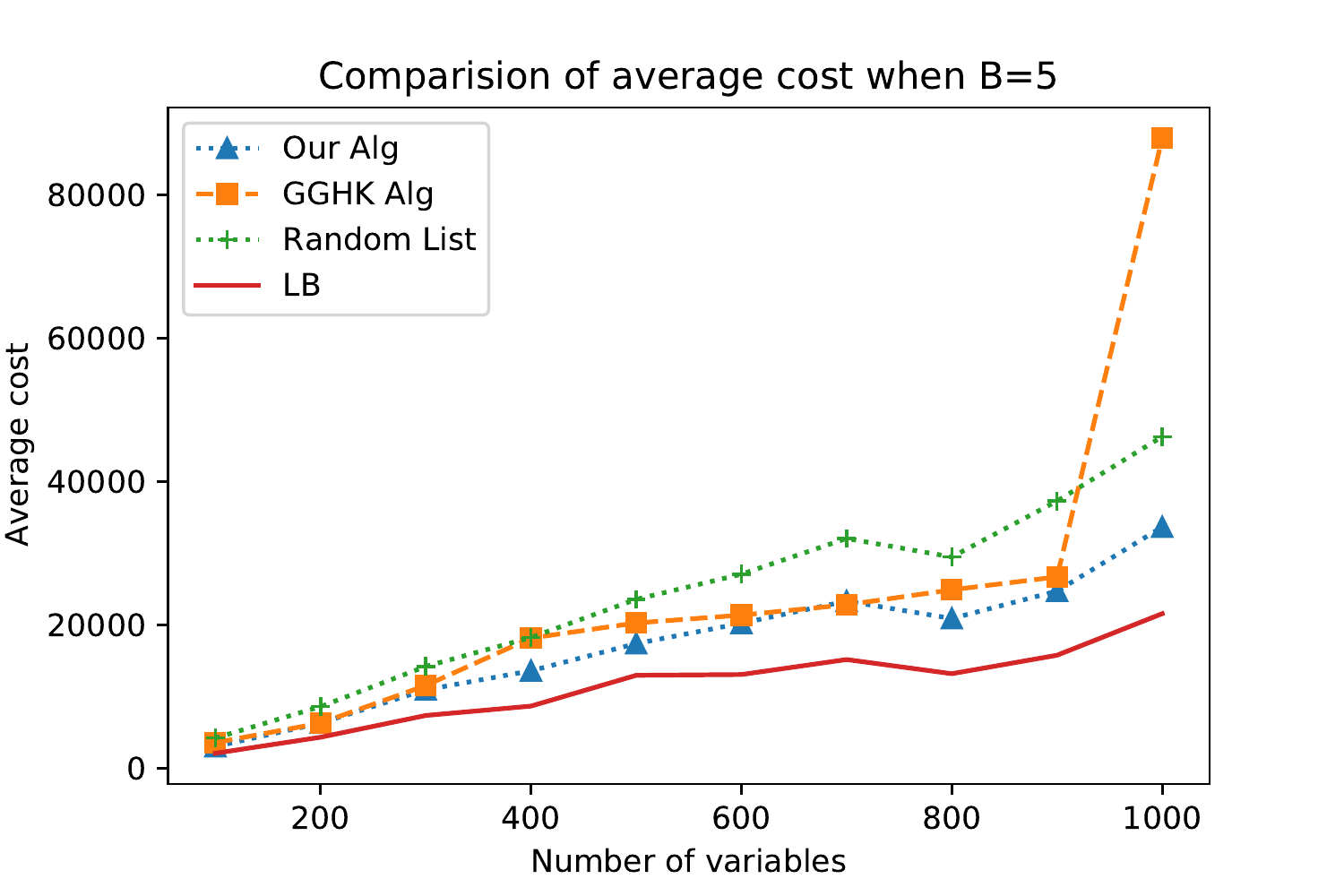}
     \end{subfigure}
     \quad
     \begin{subfigure}[b]{0.28\textwidth}
         \centering
         \includegraphics[width=2in]{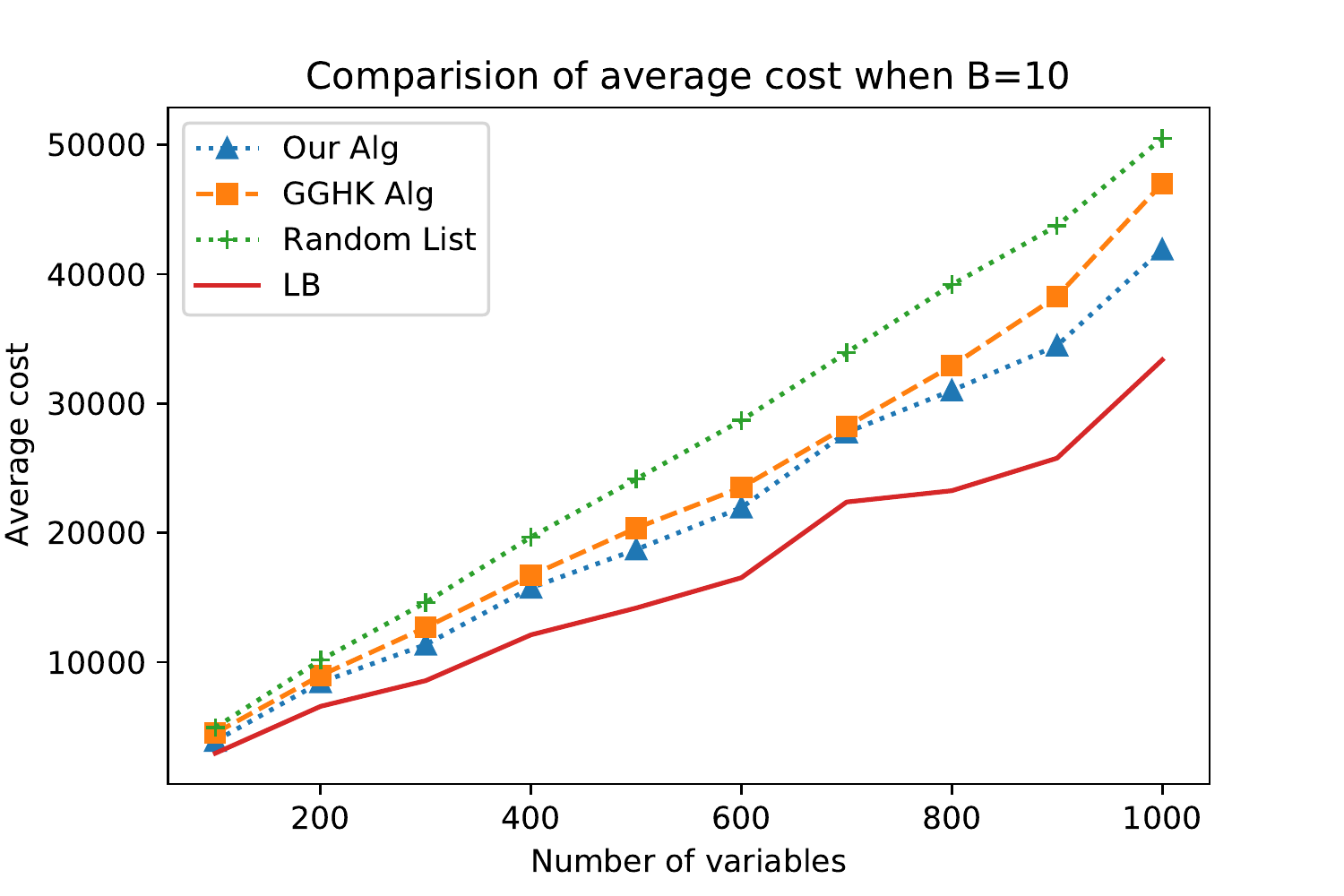}
     \end{subfigure}
     \quad
     \begin{subfigure}[b]{0.28\textwidth}
         \centering
         \includegraphics[width=2in]{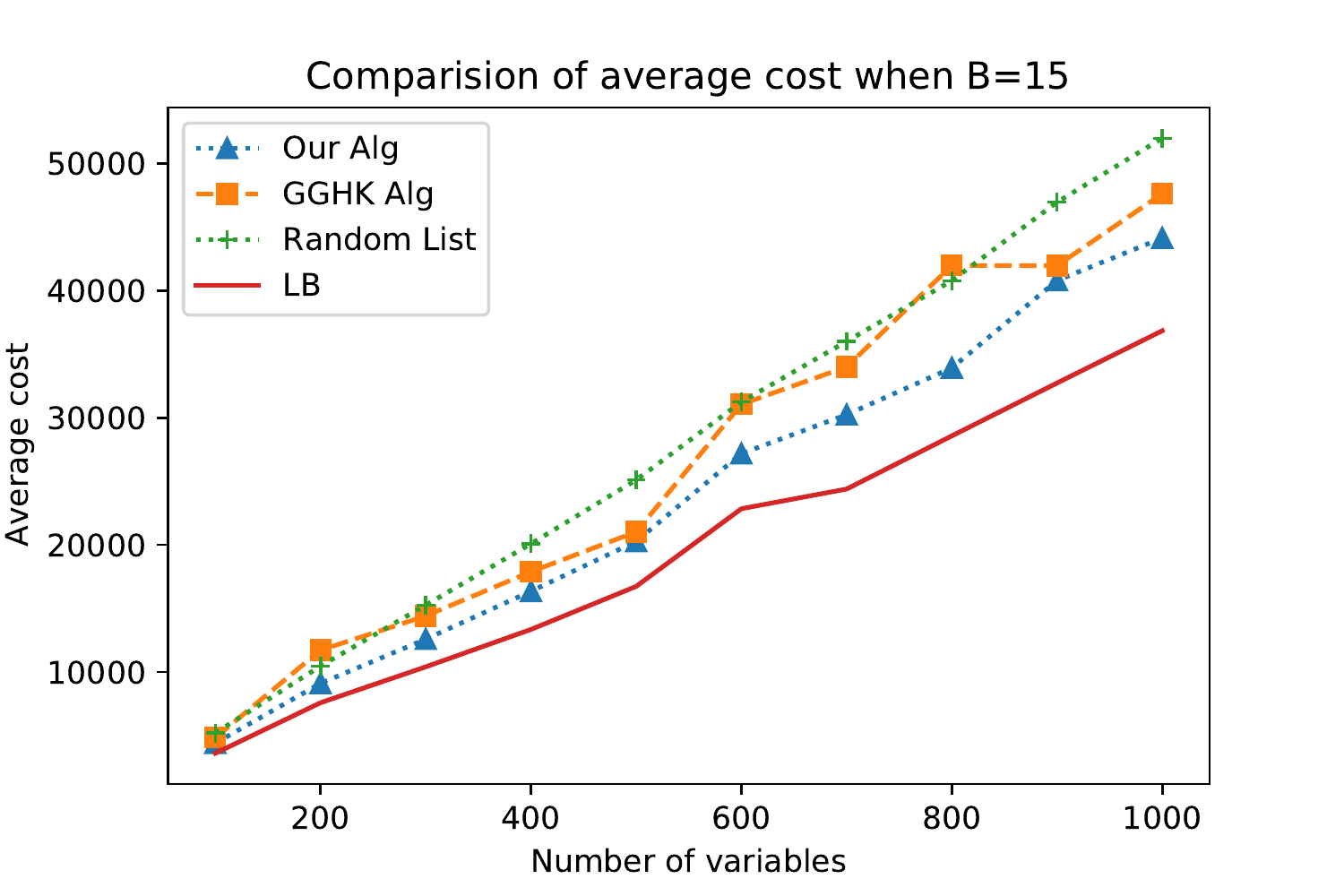}
     \end{subfigure}
     \caption{Cost results  for (weighted) $\mathtt{SSClass}$ }
        \label{fig:plots-syn-weighted-cost}
\end{figure}

\begin{figure}[h!]
     \centering
     \begin{subfigure}[b]{0.31\textwidth}
         \centering
         \includegraphics[width=2in]{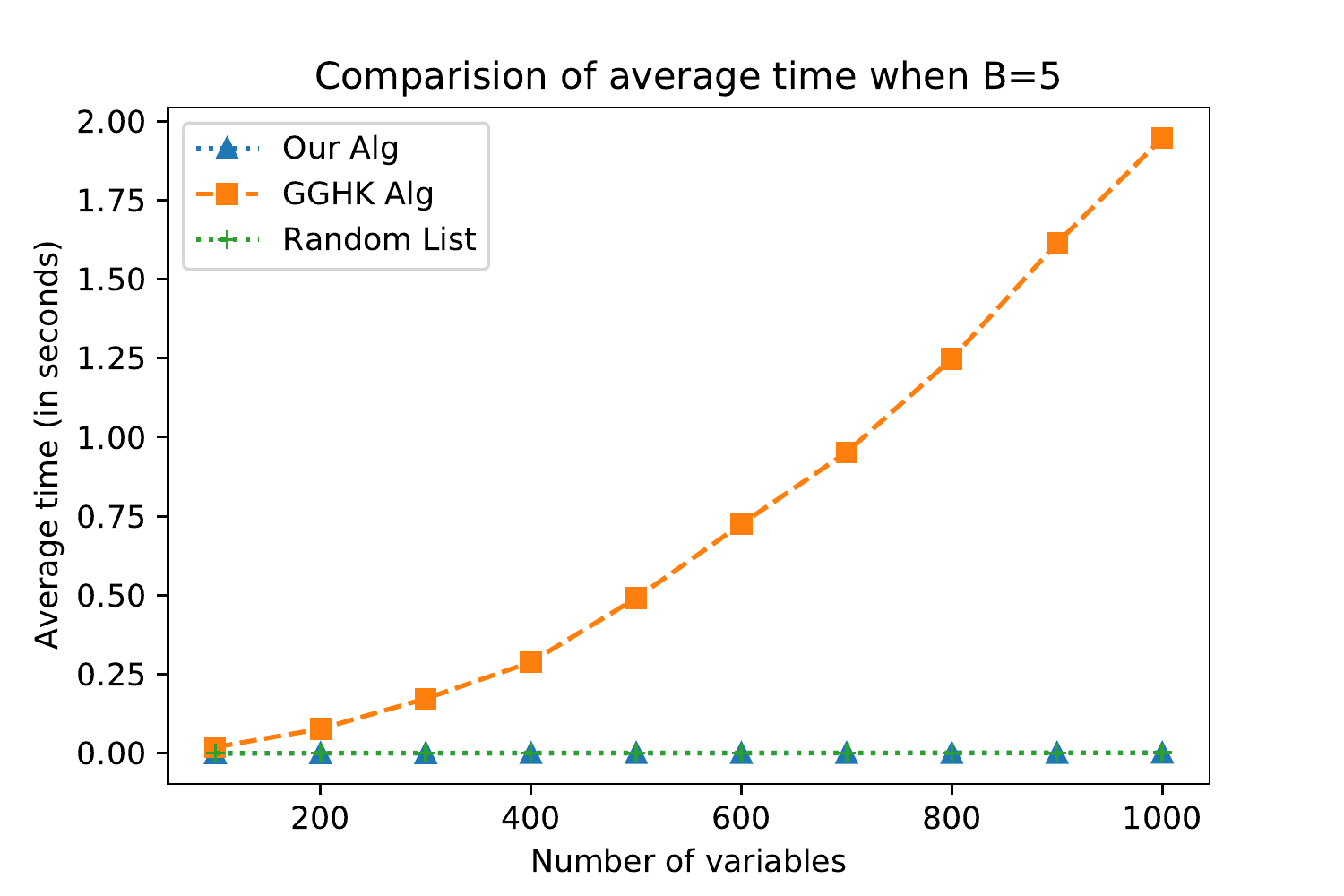}
     \end{subfigure}
     \quad
     \begin{subfigure}[b]{0.31\textwidth}
         \centering
         \includegraphics[width=2in]{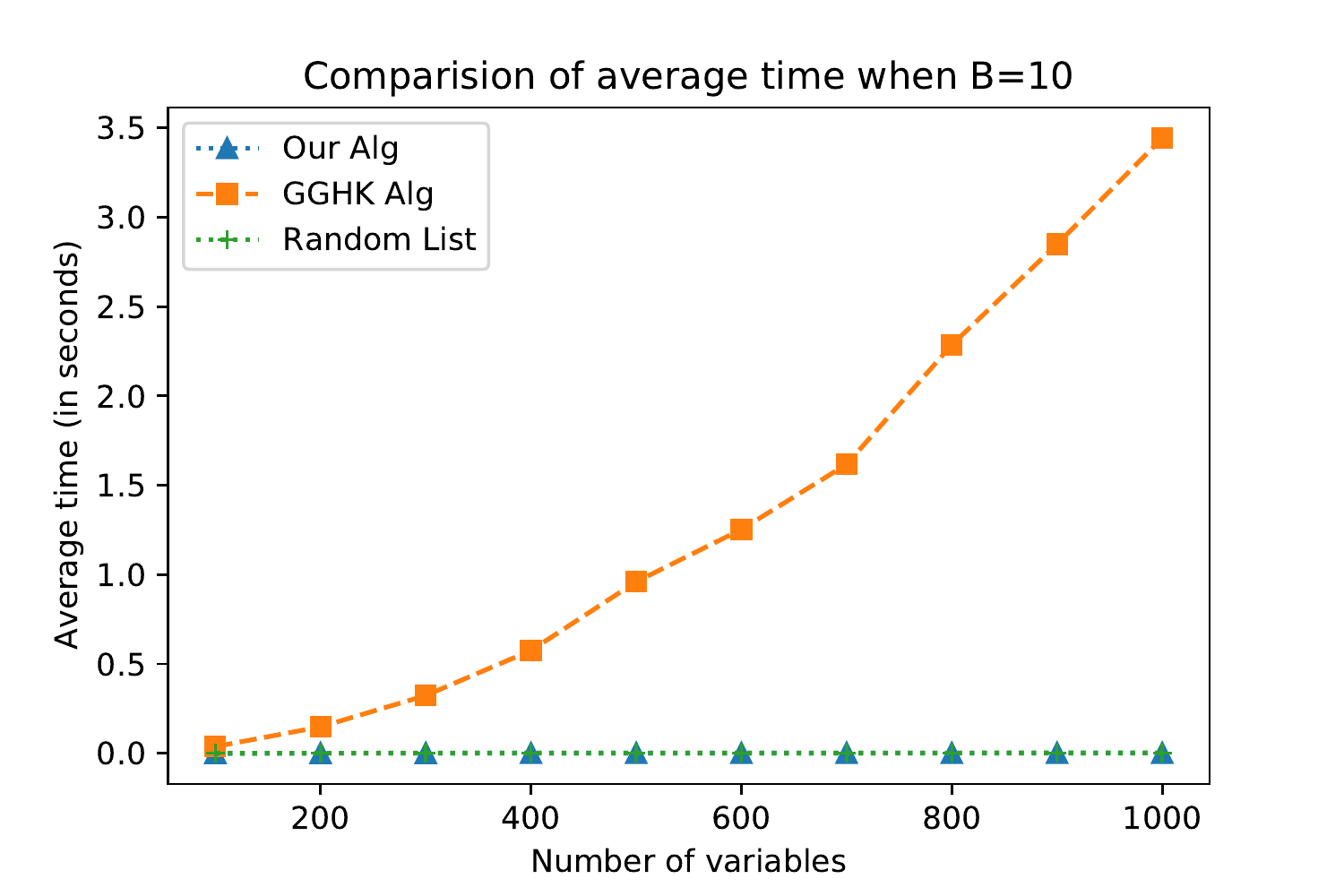}
     \end{subfigure}
     \quad
     \begin{subfigure}[b]{0.31\textwidth}
         \centering
         \includegraphics[width=2in]{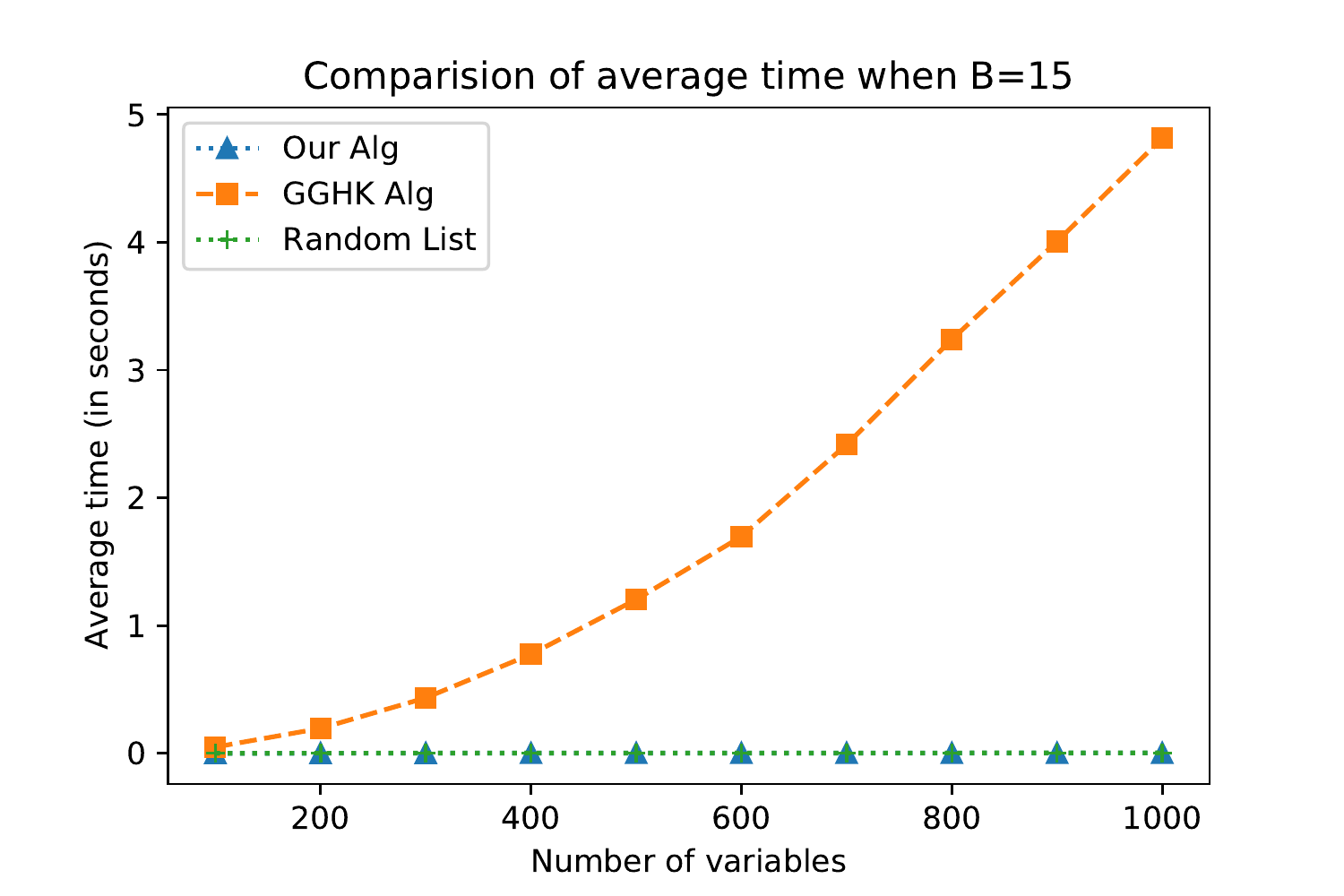}
     \end{subfigure}
     \caption{Runtime results for (weighted) $\mathtt{SSClass}$ }
        \label{fig:plots-syn-weighted-time}
\end{figure}

For each instance type (\she, Unweighted $\mathtt{SSClass}$ and  $\mathtt{SSClass}$), the plots in Figures~\ref{fig:plots-syn-LTFs}-\ref{fig:plots-syn-weighted-time} show the averages (for both cost and runtime) against the number of variables $n$. Note that each point in these plots corresponds to an average over (i) the 10 instances of its type and (ii) the 50 sampled realizations for each instance.

In \Cref{table:avg-ratios}, we report the average performance ratio (cost of the algorithm divided by  the information-theoretic lower bound) of the various algorithms. For each instance type  (\she, Unweighted $\mathtt{SSClass}$ and  $\mathtt{SSClass}$), we report the performance ratio averaged over {\em all} values of $n$ (10 choices) and all instances (10 each). Values closer to $1$ demonstrate better performance.  More detailed tables, with the average performance ratio for each value of $n$, are in  Appendix~\ref{app:tables}.

\begin{table}[h!]
\centering
{\renewcommand{\arraystretch}{0.8}%
\begin{tabular}{lccc}
\hline
Instance Type& Our Alg. & GGHK Alg. & Random List \\
\hline
Unweighted $\mathtt{SSClass}$, $B=5$ & $1.50$ & $1.48$ & $1.80$\\
Unweighted $\mathtt{SSClass}$, $B=10$ & $1.25$ & $1.24$ & $1.33$ \\
Unweighted $\mathtt{SSClass}$, $B=15$ & $1.13$ & $1.13$ & $1.19$ \\
\hline
$\mathtt{SSClass}$, $B=5$ & $1.59$ & $1.94$ & $2.43$  \\
$\mathtt{SSClass}$, $B=10$ & $1.34$ & $1.45$ & $1.73$ \\
$\mathtt{SSClass}$, $B=15$ & $1.22$ & $1.39$ & $1.47$ \\
\end{tabular}
}

\bigskip

{\renewcommand{\arraystretch}{0.8}%
\begin{tabular}{lccc}
\hline 
Instance Type & Our Alg. & DHK Alg. & Random List \\ \hline
\she & $2.18$ & $1.74$ & $5.63$ \\
\end{tabular}
}
\caption{Average performance ratios relative to the lower bound.}\label{table:avg-ratios}
\end{table}

\paragraph{Stochastic Halfspace Evaluation.} To generate an instance of \she, we set $w_i \in [10]$ uniformly for all $i \in [n]$ and select a cutoff value uniformly   in the score interval. 
We plot the results in Figure~\ref{fig:plots-syn-LTFs}. Our cost is about $20\%$ more than that of \cite{DeshpandeHK16}, but our runtime is over $100\times$ faster for large instances. 

\vspace{-8mm}

\paragraph{Unweighted Stochastic Score Classification.}
To generate instances of unweighted $\mathtt{SSClass}$, we set $w_i = 1$ for all $i \in [n]$. We choose $B \in \{5, 10, 15\}$, and then select the cutoffs (based on the value of $B$) uniformly at random in the score interval. We test our non-adaptive algorithm against the non-adaptive algorithm of \cite{GkenosisGHK18} and a random query order. Since, all the algorithms are non-adaptive, there is no difference in their running time. So, we focus on the average cost comparison among the algorithms. We observe that the average cost incurred by our non-adaptive algorithm is comparable to that of the non-adaptive algorithm of \cite{GkenosisGHK18}, and both algorithms outperform a random query order. We plot the results in Figure~\ref{fig:plots-syn-unweighted} 

\vspace{-8mm}

\paragraph{(Weighted) Stochastic Score Classification.}
We test on  $\mathtt{SSClass}$ instances with $B \in \{5, 10, 15\}$. The cutoff values $\alpha_j$ are selected uniformly at random in the score interval.  We plot results in Figures~\ref{fig:plots-syn-weighted-cost} and \ref{fig:plots-syn-weighted-time}. In all cases, we observe that our non-adaptive algorithm beats the adaptive algorithm in both query cost and runtime; for e.g., when $B=10$ and $n=900$ our cost is $10\%$ less and  our runtime is about $100\times$ faster.

\vspace{-10mm}

\vspace{-10mm}\section{Conclusions}
In this paper, we obtained the first constant-factor approximation algorithm for stochastic score classification, which   significantly  
generalizes sequential testing algorithms for  series systems, $k$-of-$n$ functions and halfspaces. Our algorithm is non-adaptive and very efficient to implement. We also evaluated our algorithm  empirically and compared it to previous approaches and an information-theoretic lower bound: in most cases, our algorithm provides an improvement in solution cost or runtime. Our approach also extends to other sequential testing problems such as intersection of $d$ halfspaces, and to the setting of batched tests.

 \section*{Acknowledgements}
 A preliminary version of this paper (with the same title) appeared in the conference on Integer Programming and Combinatorial Optimization (IPCO), 2022.

\bibliographystyle{alpha}
\bibliography{main}

\appendix
\section{Handling Negative Weights}\label{app:positive-wt}
We now show how our results can be extended to the case where the weights $w_i$ may be positive or negative. 
For the stochastic score classification problem ($\mathtt{SSClass}$), we provide a reduction from any instance (with aribtrary weights) to one with positive weights. Let ${\cal I}$ be any instance of $\mathtt{SSClass}$, and let $P\sse [n]$ (resp. $N\sse [n]$) denote the items with positive (resp. negative) weight. Note that we can re-write the total weight as follows:
\begin{equation}\label{eq:app-neg-wt}
    w(X) = \sum_{i=1}^n w_i X_i = \sum_{i\in P}  w_i X_i - \sum_{i\in N} |w_i|X_i= - \sum_{i\in N} |w_i| \,+\,  \sum_{i\in P}  w_i X_i + \sum_{i\in N} |w_i|(1-X_i).
\end{equation}
Consider now a new $\mathtt{SSClass}$ instance with the $n$ variables $X'_i= X_i$ for ${i\in P}$ and $X'_i=1-X_i$ for ${i\in N}$. The probabilities are $p'_i=p_i$ for ${i\in P}$ and $p'_i=1-p_i$ for ${i\in N}$. The weights  are $w'_i=w_i$ for ${i\in P}$ and $w'_i=-w_i$ for ${i\in N}$; note that all weights are now positive.  The costs remain the same as before. For any realization $X$, the new weight  is $\sum_{i=1}^n w'_i X'_i = \sum_{i\in N} |w_i| + w(X)$, using \eqref{eq:app-neg-wt}. Finally the class boundaries for the new instance are $\alpha'_j=\alpha_j + \sum_{i\in N} |w_i|$. It is easy to check that a realization $X'$ in the new instance has class $j$ if and only if the corresponding realization $X$ has class $j$ in the original instance ${\cal I}$.

\paragraph{Explainable $d$-halfspace evaluation.}  To handle negative weights in   \exsfe, we need to apply the above idea within the algorithm (it is not a black-box reduction to positive weights). At a high level, we use the fact that the algorithm simply interleaves the separate lists for each halfspace.  

Consider any halfpsace $h_k(X) =\langle \sum_{i=1}^n w_{ki} X_i\ge \alpha_k\rangle$ with both positive and negative weights. Let $P_k\sse[n]$ and $N_k=[n]\setminus P_k$ denote the items with positive and negative weight respectively (for halfspace $k$). We then rephrase $h_k(X)$ as the following halfspace $h'_k(X)$.
\begin{equation*} 
     \sum_{i\in P_k}  w_{ki} X_i + \sum_{i\in N_k} |w_{ki}|(1-X_i)\ge \alpha_k + \sum_{i\in N_k} |w_{ik}|.  
\end{equation*}
Correspondingly, we define two {\em non-negative} rewards $R_{k,0}$ and $R_{k,1}$ (for  halfspace $k$) as follows.
\begin{equation*}
    R_{k,0}(i) = \begin{cases} w_{ki} (1 - X_i) & \text{ if } i \in P_k \\ -w_{ki} X_i & \text{ if } i \in N_k\end{cases}, \qquad R_{k,1}(i) = \begin{cases} w_{ki} X_i & \text{ if } i \in P_k \\ -w_{ki} (1- X_i) & \text{ if } i \in N_k\end{cases}
\end{equation*}
Note that after probing a subset $S\sse[n]$, the total $R_{k,0}$ (resp. $R_{k,1}$) reward corresponds to a lower (resp. upper) bound on the left-hand-side of the modified halfspace $h'_k(X)$, which allows us to evaluate the original halfspace $h_k(X)$.  The rest of the algorithm  remains the same as in \S\ref{sec:general}. The analysis also remains the same. 

\section{Proof of Theorem~\ref{thm:knap}}
Recall the setting of Theorem~\ref{thm:knap}. We are given a set $T$ of items with non-negative costs $\{c_k\}$ and rewards $\{r_k\}$, along with a budget $D$. The goal is to select a subset of items of total cost at most $D$ that maximizes the total reward. The following is a natural LP relaxation of the knapsack problem where the objective is expressed as a function of the budget $D$.
$$g(D) = \max \left\{ \sum_{k\in T} r_k\cdot x_k \,\, \big| \,\,\sum_{k\in T} c_k\cdot x_k \le D, \, x\in [0,1]^T\right\},\qquad \forall D\ge 0.$$ 
The following algorithm $\A_{\rv{KS}}$ solves the {\em fractional} knapsack problem and also obtains an approximate integral solution. 
Assume that the items are ordered so that $\frac{r_1}{c_1}\ge \frac{r_2}{c_2}\ge \cdots$. Let $t$ index the first item (if any) so that $\sum_{k=1}^t c_k\ge D$. Let $\psi:=\frac{1}{c_t}(D-\sum_{k=1}^{t-1} c_k)$ which lies in $(0,1]$. Define
$$x_k=\left\{ \begin{array}{ll}
1 & \mbox{ if } k\le t-1\\
\psi & \mbox{ if } k=t
\end{array}\right..$$
Return $x$ as the optimal fractional solution and $Q=\{1,\cdots, t\}$ as an integer solution. We restate Theorem~\ref{thm:knap} for completeness.
\begin{theorem}
Consider algorithm $\A_{\rv{KS}}$ on any instance of the knapsack problem with budget $D$. 
\begin{enumerate}
\item $g(D) = \sum_{k=1}^{t-1} r_k +\psi\cdot r_t = \langle r,
  x\rangle$ and so $x$ is an optimal LP solution. 
\item The derivative $g'(D) = \frac{r_t}{c_t}$.
\item Solution $Q$ has cost $c(Q)\le D+c_{max}$ and reward $r(Q)\ge g(D)$. 
\item $g(D)$ is a concave function of $D$.
\end{enumerate}
\end{theorem}
\proof{Proof.}
Let $x^*$ be an optimal LP solution. If $x^* = x$, we are done. Suppose that $x^* \neq x$, and assume without loss of generality that $\sum_{k=1}^nc_k \cdot x_k^* = D$; else we can obtain a greater reward by raising $x^*$. Let $j$ be the smallest index such that $x_j^* < x_j$: note that $j \leq t$ is well defined by the definition of $x$. Let $h$ be the largest index  with $x^*_h>x_h$: note that such an index must exist as $\sum_{k=1}^nc_k \cdot x_k^* =\sum_{k=1}^nc_k \cdot x_k $. Moreover, $h\ge t$ by definition of our solution $x$. As $j\ne h$, we have $j<h$ from above. Define a new solution
$$x_k'=\left\{ \begin{array}{ll}
x^*_k & \mbox{ if } k\ne j,h\\
x^*_j+\delta & \mbox{ if } k = j\\
x^*_h-\frac{c_j}{c_h}\delta & \mbox{ if } k = h
\end{array}\right..$$
Above, $\delta=\min\{1-x^*_j , \frac{c_h}{c_j} x^*_h\}>0$. 
Intuitively, we are redistributing cost from item $h$ to $j$. Note that the cost of the new solution $\sum_{k=1}^nc_k \cdot x_k' = \sum_{k=1}^nc_k \cdot x_k^*=D$. Moreover, the reward 
$$\sum_{k=1}^n r_k\cdot x'_k = \sum_{k=1}^n r_k\cdot x^*_k + \delta r_j - \delta\frac{c_j}{c_h} r_h \ge \sum_{k=1}^n r_k\cdot x^*_k,$$
where we used $\frac{r_j}{c_j} \geq \frac{r_h}{c_h}$ which follows from $j<h$ and the ordering of items. 
Finally, by choice of $\delta$, either $x'_j=1$ or $x'_h=0$. It follows that $x'$ is also an optimal LP solution. Repeating this process, we obtain that $x$ is also an optimal LP solution. This completes the proof of property (1). 

For property (2), observe that $$ g'(D) = \lim_{\epsilon \to 0}\frac{g(D+\epsilon) - g(D)}{\epsilon} = \lim_{\epsilon \to 0} \frac{g(D) + \epsilon \cdot \frac{r_t}{c_t} - g(D)}{\epsilon} = \frac{r_t}{c_t}, $$ as desired. 

Since $\psi \in (0, 1]$, we have $c(Q) = D + (1 - \psi) \cdot c_t \leq D + c_{\max}$ and $r(Q) = (1 - \psi) \cdot r_t + g(D) \geq g(D)$, proving property (3).

Finally, using property (2) and the non-increasing $\frac{r_k}{c_k}$ order of the items, it follows that $g(D)$ is a concave function.  This proves property (4).
\endproof

\section{An Information-Theoretic Lower Bound for $\mathtt{SSClass}$}\label{sec:comp-lb}
Here, we present an information theoretic lower bound for $\mathtt{SSClass}$. Recall that an instance of $\mathtt{SSClass}$ consists of $n$ independent Bernoulli random variables $X = X_1, \ldots, X_n$, where variable $X_i$ is $1$ with probability $p_i$, and its realization can be probed at cost $c_i \in \R_+$. The score of the outcome $\bX = (X_1, \ldots, X_n)$ is
$r(\bX) = \sum_{i=1}^n w_i X_i$ where $w_i\in \ZZ_+$ for all $i \in [n]$. Additionally, we are given $B+1$ thresholds $\alpha_1, \ldots, \alpha_{B+1}$ which define intervals $I_1, \ldots, I_B$ where $I_j = \{\alpha_j, \ldots, \alpha_{j+1}-1\}$. These intervals define a \emph{score
classification function} $h: \{0, 1\}^n \to \{1, \ldots, B\}$;
$h(\bX)=j$ precisely when $r(\bX) \in I_j$. The goal is to
determine $h(\bX)$ at minimum expected cost.

Let $\widehat{X} = (\widehat{X}_1, \ldots, \widehat{X}_n)$ correspond to a realization of the variables $\bX$. Furthermore, suppose that $h(\widehat{X})=j$; that is, under realization $\widehat{X}$, the score $h(\widehat{X})$ lies in $I_j$. Let $S \sse [n]$ correspond to the set of probed variables. Recall that $S$ is a feasible solution for $\mathtt{SSClass}$ under realization $\widehat{X}$ when the following conditions on the $R_0$ and $R_1$ rewards of $S$ hold: $\sum_{i \in S}R_0(i) = \sum_{i \in S} w_i \cdot (1 - \widehat{X}_i) \geq \beta_j^0$ and $\sum_{i \in S}R_1(i) = \sum_{i \in S}w_i \cdot \widehat{X}_i \geq \beta_j^1$ where $\beta_j^0 = W - \alpha_{j+1} + 1$ and $\beta_j^1 = \alpha_j$. Thus, the following integer program computes a lower bound on the probing cost needed to conclude that $h(\widehat{X})=j$. 
\begin{align}
    \text{minimize} \qquad & \sum_{i=1}^n c_i \cdot z_i \notag \\
    \text{subject to} \qquad & \sum_{i=1}^n w_i \widehat{X}_i \cdot z_i \geq \beta_j^1 \label{eq:lb-ip}\\
    & \sum_{i=1}^nw_i (1-\widehat{X}_i) \cdot z_i \geq \beta_j^0 \notag \\
    & z_i \in \{0, 1\} \quad i \in [n] \notag
\end{align}

where $z_i$, for $i \in [n]$, is a binary variable denoting whether $X_i$ is probed. Let $\widehat{\mathtt{LB}}$ denote the optimal value of \eqref{eq:lb-ip}. Then, $\mathtt{LB} = \E[\widehat{\mathtt{LB}}]$ is an information-theoretic lower bound for the given $\mathtt{SSClass}$ instance. 
We note that \eqref{eq:lb-ip} only provides a lower bound on the probing cost for realization $\widehat{X}$, and is not a formulation for the given $\mathtt{SSClass}$ instance.

\section{Additional Tables}\label{app:tables}
Here we give more detailed tables for  the average performance ratio (cost of the algorithm divided by  the information-theoretic lower bound) of the various algorithms. For each instance type  (\she, Unweighted $\mathtt{SSClass}$ and  $\mathtt{SSClass}$) and each value of $n$, we report the performance ratio averaged over all instances (10 each). Values closer to $1$ demonstrate better performance.

\begin{table}[h!]
\centering
{\renewcommand{\arraystretch}{0.8}%
\begin{tabular}{lccc}
\hline
$n$& Our Alg. & GGHK Alg. & Random List \\
\hline
$100$ & $1.46$ & $1.45$ & $1.73$ \\
$200$ & $1.35$	& $1.34$ &	$1.54$ \\
$300$ & $1.46$	&$1.45$	&$1.76$ \\
$400$ & $1.52$	&$1.50$	&$1.81$ \\
$500$ & $1.67$	&$1.65$	&$2.24$ \\
$600$ & $1.50$	&$1.48$	&$1.77$ \\
$700$ & $1.52$	&$1.51$	&$1.80$ \\
$800$ & $1.52$	&$1.50$	&$1.79$ \\
$900$ & $1.52$	&$1.51$	&$1.72$ \\
$1000$ & $1.45$	&$1.44$&	$1.78$ \\
\hline
\end{tabular}
}
\caption{Unweighted $\mathtt{SSClass}$, $B=5$}
\end{table}

\begin{table}[h!]
\centering
{\renewcommand{\arraystretch}{0.8}%
\begin{tabular}{lccc}
\hline
$n$& Our Alg. & GGHK Alg. & Random List \\
\hline
100  & 1.31 & 1.30 & 1.39 \\
200 & 1.26 & 1.25 & 1.32 \\
300 & 1.15 & 1.15 & 1.26 \\
400 & 1.27 & 1.26 & 1.40 \\
500 & 1.32 & 1.31 & 1.40 \\
600 & 1.21 & 1.20 & 1.28 \\
700 & 1.21 & 1.20 & 1.36 \\
800 & 1.20 & 1.20 & 1.28 \\
900 & 1.24 & 1.23 & 1.29 \\
1000 & 1.29 & 1.29 & 1.36 \\                   
\hline
\end{tabular}
}
\caption{Unweighted $\mathtt{SSClass}$, $B=10$}
\end{table}

\begin{table}[h!]
\centering
{\renewcommand{\arraystretch}{0.8}%
\begin{tabular}{lccc}
\hline
$n$& Our Alg. & GGHK Alg. & Random List \\
\hline
100 & 1.12 & 1.11 & 1.15 \\
200 & 1.16 & 1.16 & 1.20 \\
300 & 1.21 & 1.20 & 1.34 \\
400 & 1.09 & 1.09 & 1.14 \\
500 & 1.12 & 1.12 & 1.16 \\
600 & 1.13 & 1.13 & 1.17 \\
700 & 1.15 & 1.15 & 1.25 \\
800 & 1.12 & 1.12 & 1.14 \\
900 & 1.11 & 1.11 & 1.16 \\
1000 & 1.10 & 1.10 & 1.15\\                   
\hline
\end{tabular}
}
\caption{Unweighted $\mathtt{SSClass}$, $B=15$}
\end{table}

\begin{table}[h!]
\centering
{\renewcommand{\arraystretch}{0.8}%
\begin{tabular}{lccc}
\hline
$n$& Our Alg. & GGHK Alg. & Random List \\
\hline
100 & 1.56 & 1.97 & 2.47 \\
200 & 1.55 & 1.46 & 2.42 \\
300 & 1.59 & 1.69 & 2.19 \\
400 & 1.61 & 2.07 & 2.31 \\
500 & 1.38 & 1.65 & 1.95 \\
600 & 1.63 & 1.72 & 2.41 \\
700 & 1.63 & 1.55 & 2.35 \\
800 & 1.74 & 1.87 & 3.39 \\
900 & 1.64 & 1.78 & 2.54 \\
1000 & 1.60 & 3.69 & 2.26\\                   
\hline
\end{tabular}
}
\caption{$\mathtt{SSClass}$, $B=5$}
\end{table}

\begin{table}[h!]
\centering
{\renewcommand{\arraystretch}{0.8}%
\begin{tabular}{lccc}
\hline
$n$& Our Alg. & GGHK Alg. & Random List \\
\hline
100 &1.35 & 1.55 & 1.73 \\
200 & 1.29 & 1.37 & 1.57 \\
300 & 1.34 & 1.53 & 1.79 \\
400 & 1.35 & 1.44 & 1.76 \\
500 & 1.35 & 1.46 & 1.77 \\
600 & 1.36 & 1.44 & 1.83 \\
700 & 1.30 & 1.28 & 1.71 \\
800 & 1.37 & 1.46 & 1.79 \\
900 & 1.39 & 1.54 & 1.82 \\
1000 & 1.26 & 1.46 & 1.58\\                   
\hline
\end{tabular}
}
\caption{$\mathtt{SSClass}$, $B=10$}
\end{table}

\begin{table}[h!]
\centering
{\renewcommand{\arraystretch}{0.8}%
\begin{tabular}{lccc}
\hline
$n$& Our Alg. & GGHK Alg. & Random List \\
\hline
100 & 1.23 & 1.36 &1.47 \\
200 & 1.20 & 1.54 & 1.39 \\
300 & 1.22 & 1.43 & 1.52 \\
400 & 1.25 & 1.39 & 1.56 \\
500 & 1.22 & 1.26 & 1.53 \\
600 & 1.19 & 1.38 & 1.38 \\
700 & 1.25 & 1.41 & 1.50 \\
800 & 1.19 & 1.51 & 1.46 \\
900 & 1.26 & 1.29 & 1.46 \\
1000 & 1.23 & 1.31 & 1.46\\                   
\hline
\end{tabular}
}
\caption{$\mathtt{SSClass}$, $B=15$}
\end{table}

\begin{table}[h!]
\centering
{\renewcommand{\arraystretch}{0.8}%
\begin{tabular}{lccc}
\hline
$n$& Our Alg. & DHK Alg. & Random List \\
\hline
100 & 2.21 & 1.85 & 4.14 \\
200 & 2.14 & 1.61 & 6.28 \\
300 & 2.36 & 1.66 & 6.79 \\
400 & 2.28 & 1.66 & 7.15 \\
500 & 2.26 & 1.75 & 7.58 \\
600 & 2.13 & 1.74 & 5.70 \\
700 & 2.12 & 1.81 & 4.80 \\
800 & 2.12 & 1.76 & 5.13 \\
900 & 2.10 & 1.77 & 4.75 \\
1000 & 2.11 & 1.85 & 3.96\\                   
\hline
\end{tabular}
}
\caption{Stochastic Halfspace Evaluation (\she)}
\end{table}

\end{document}